\documentclass[final]{elsarticle}
\usepackage{lineno,hyperref}
\usepackage{pgf,tikz,pgfplots}
\usepackage{mathrsfs}
\usepackage{amsmath}
\usepackage{tikz}
\usepackage{mathrsfs}
\usetikzlibrary{positioning}
\usetikzlibrary{arrows}
\usepackage{subfig}
\usepackage{amsmath}
\usepackage{tabularx}
\usepackage{algpseudocode,algorithm}
\usepackage{algorithm}
\usepackage{commath}
\usepackage{listings}
\usepackage{graphicx}
\usepackage{comment}
\usepackage{fp}

\modulolinenumbers[5]
\journal{Computer Physics Communications}

\begin{document}

\begin{frontmatter}

\title{PittPack: An Open-Source Poisson's Equation Solver for Extreme-Scale Computing with Accelerators}
	\author{Jaber J. Hasbestan\fnref{myfootnote}}
	\fntext[myfootnote]{Postdoctoral Research Associate, jaber@pitt.edu}
	\author{Inanc Senocak\fnref{myfootnote1}}
	\fntext[myfootnote1]{Associate Professor, senocak@pitt.edu}
	\address{Department of Mechanical Engineering and Materials Science,\\ University of Pittsburgh, Pittsburgh, PA 15261, USA}

\begin{abstract}
We present a parallel implementation of a direct solver for the Poisson's equation on extreme-scale supercomputers with accelerators. We introduce a chunked-pencil decomposition as the domain-decomposition strategy to distribute work among processing elements to achieve superior scalability at large number of accelerators.
Chunked-pencil decomposition enables overlapping nodal communication and data transfer between the central processing units (CPUs) and the graphics processing units (GPUs). Second, it improves data locality by keeping neighboring elements in adjacent memory locations. 
Third, it allows usage of shared-memory for certain segments of the algorithm when possible, and last but not least, it enables contiguous message transfer among the nodes. 
Two different communication patterns are designed. The fist pattern aims to fully overlap the communication with data transfer and designed for speedup of overall turnaround time, whereas the second method concentrates on low memory usage and is more network friendly for computations at extreme scale. To ensure software portability, we interleave \texttt{OpenACC} with \texttt{MPI} in the software. The numerical solution and its formal second order of accuracy is verified using method of manufactured solutions for various combinations of boundary conditions.
Weak scaling analysis is performed using up to 1.1 trillion Cartesian mesh points using 16384 GPUs on a petascale leadership class supercomputer.

\end{abstract}

\end{frontmatter}


\section{Introduction}
Poisson's equation is an elliptic partial differential equation. It is used to represent several physical phenomena such as gravitational, electrostatic fields and enforce conservation of mass principle in the numerical solution of incompressible fluid flow equations. 
Poisson's equation for a three-dimensional domain, $\Omega$, with the appropriate boundary conditions defined on $\partial \Omega$ can be written in the following compact form:

\begin{equation}
\begin{split}
\mathcal{L} u=f \hspace{0.2cm} & in \hspace{0.2cm} \Omega,  \\
\mathcal{G} u=0 \hspace{0.2cm} & on \hspace{0.2cm} \partial \Omega,
\end{split}
\label{eq:poisson}
\end{equation}

\noindent
where $\mathcal{L}$ is the Laplacian operator, and $f$ is a source term and $\mathcal{G}$ specifies the boundary conditions on the solution variable $u$. 
Equation (\ref{eq:poisson}) can be solved using various numerical methods such as finite-volume, finite-difference, or finite element formulations \cite{karniadakis2013spectral}.
Regardless of the selected approach, the final stage of the solution requires the solution of
a linear system of equations which can be written in a compact form as a matrix-equation as follows, 
\begin{equation}
\textit{A}\textbf{x}=\textbf{b},    
\end{equation}
\noindent
where $A$ is the coefficient matrix and \textbf{x} is the solution vector and \textbf{b} is the vector of constants. The efficiency of a numerical method to solve the above system of equations highly depends on the 
properties of matrix $A$, specifically its bandwidth and condition number. Discretization scheme, desired order of accuracy in the solution and the underlying computational grid affect the properties of the coefficient matrix. The solution of the Poisson's equation can be the most time-consuming part of an algorithm. Therefore, fast, massively-parallel methods are highly prized in practice.


Both the condition number and the bandwidth of the matrix $A$ are important factors when it comes to increasing the efficiency of the numerical solution. Solution methodologies that rely on iterative techniques, such as the conjugate gradient (CG) \cite{shewchuk1994introduction} or the generalized minimal residual (GMRES) method \cite{saad1986gmres,saad2003iterative}, focus on improving the search directions and condition number by adopting improved preconditioners, whereas direct solution techniques are concerned with bandwidth reduction, but this strategy is not generally applicable to nonlinear PDEs. Matrix reordering techniques such as Cuthill-Mckee \cite{cuthill1969reducing} are commonly used in this realm with the idea of generating less non-zero element in the factorization process.
Fortunately, Poisson's equation is a linear PDE. Bandwidth reduction can then be achieved through separation of variables in the continuous form of the linear PDEs \cite{kreyszig2010advanced}. To this end, fast Fourier transform (FFT) \cite{hockney1965fast} was utilized. 

Application of FFT to solve Poisson's equation was first realized by Hockney \cite{hockney1965fast}
and is subsequently extended to address a general class of separable elliptic equations by Swarztrauber \cite{swarztrauber1974direct} and to three dimensions by Wilhelmson \cite{wilhelmson1977direct}. Later, Swarztrauber \cite{swarztrauber1977methods} introduced the Fourier and Cyclic Reduction {(FACR)} algorithm, which is a combination of cyclic reduction and Fourier transform and its asymptotic operations count is less than either the Fourier transform or the cyclic reduction used independently. Hence, FACR is computationally efficient while using the same amount of memory. 
The idea of separation of variables and its potential for parallelism is not new and was recognized by Buzbee \cite{buzbee1973fast} in 1973. Considering that the computer hardware is constantly evolving, parallel implementation of popular algorithms needs to be revisited to enable efficient utilization of the ever increasing computational capability.  

Fortunately, there are FFT libraries developed for computing on CPUs and GPUs.
\texttt{FFTW} is a popular package developed by Frigo et al. \cite{FFTW05} for CPUs.
It uses Cooley-Tukey algorithm \cite{cooley1965algorithm,cooley1969fast,cooley1970fast}. Recent version of \texttt{FFTW} provide some parallel computing features. There are also other libraries specialized to perform FFT in parallel. \texttt{P3DFFT} \cite{pekurovsky2012p3dfft} uses pencil decomposition and was shown to scale up to 524,288 cores. It uses \texttt{\texttt{MPI}\_Alltoallv} to perform \texttt{MPI} transformation. The package also supports hybrid programming with \texttt{\texttt{MPI}/OpenMp}.  
A \texttt{C++} version of \texttt{P3DFFT}, called \texttt{P3DFFT++}, is also available and extends the interface of \texttt{P3DFFT}. It provides the user with a choice in defining their own data layout formats beyond the predefined 2D pencil blocks. 

\texttt{2DECOMP \& FFT} library is introduced by Li et al. \cite{li20102decomp}. It is developed in Fortran and designed for applications using three-dimensional structured mesh and spatially implicit numerical algorithms. It implements a general-purpose 2D pencil decomposition for data distribution on distributed-memory platforms.
AFiD \cite{zhu2018afid} uses this \texttt{2DECOMP} library to perform the decomposition for simulation of the turbulent incompressible flows. 
Some signal processing applications require d-dimensional transforms, where $d$ can be larger than $3$. To address this need, Pipping  developed \texttt{PFFT} \cite{pippig2013pfft} which computes d-dimensional FFT with r-dimensional data decomposition by using a generalized multidimensional pencil decomposition. 

\texttt{AccFFT} \cite{gholami2015accfft} was developed based on \texttt{cuFFT} \cite{nvidia2010cufft} to calculate FFT on GPU platforms It is also an open-source software and was shown to scaling analysis for up to 4096 K20 GPUs on Titan supercomputer at the Oak Ridge National Laboratory (ORNL). It also supports CPU clusters and was tested on stampede.  




The heterogeneous architecture of modern computing clusters with accelerators adds another layer of complication to parallel software development. The architecture often dictates new strategies and algorithms in order to extract the best performance from the computing cluster.
This, in part, arises due to the difference in hardware architecture between CPU and GPU as well as the
need to migrate the data between the CPU and GPU and the desire to perform these copy operations asynchronously whenever possible. 
In other words, algorithms that thrive on CPU only might not perform well on hybrid systems. Therefore, it is vital to review algorithmic decisions before implementing a numerical techique for heteregenous systems. Naively porting a champion CPU algorithm to GPU might exhibit unsatisfactory performance behavior.

The need for efficient software to utilize GPU accelerators led to the development of several open-source libraries in \texttt{CUDA}. For example, NVIDIA offers \texttt{cuBLAS} \cite{toolkit20114} and cuFFT \cite{nvidia2010cufft}. \texttt{cuFFT} has similar functionality as \texttt{FFTW} \cite{frigo2005design} but it is primarily designed to perform FFT transforms on batches of data concurrently on a single GPU. These libraries are invaluable and used as building blocks to develop specialized and sophisticated simulation software in various engineering and scientific domains.



Almost all of the software that support 3D FFT operations can be used with minor effort to solve Poisson's equation with spectral accuracy. However, for a large number of fluid flow applications second order accuracy is sufficient because Poisson's equation is solved along with momentum and scalar transport equations that are typically discretized using second order accurate finite volume or difference schemes. It is worth mentioning that finite difference or volume type methods on Cartesian meshes map well to the memory architecture of GPUs to yield superior performance as opposed to unstructured mesh techniques. It is a well-known fact that a large portion of the simulation of incompressible flow (i.e., roughly 60\%) is spent on the solution of the Poisson's equation to enforce the continuity condition. Therefore, any improvement on this segment of the simulation drastically improves the overall simulation time. 


In the present work, we offer \texttt{PittPack}  as a massively parallel second-order accurate Poisson's equation solver designed to execute efficiently on heterogeneous computing clusters with accelerators.  \texttt{PittPack} is portable in the sense that not only it can target different GPUs from different vendors (NVIDIA, AMD, etc), but also it can be compiled to run on CPU-only clusters. Thanks to the portability of \texttt{OpenACC}.
It can take advantage of the accelerators when available, otherwise it can be used on regular CPU clusters. Switching between CPU and GPU in the code is trivial and done by setting \texttt{\_OPENACC} preprocessor variable to zero or one. 
The main algorithmic contributions of the present work are as follows:
\begin{itemize}
\item Introduction of a chunked-pencil decomposition as an efficient domain-decomposition strategy for heterogeneous clusters with accelerators 
\item Design of a custom pairwise exchange algorithm to overlap communication and data transfer between the host and the device, we extend the algorithm used in collective communication in \texttt{MPI} libraries  \cite{thakur2005optimization} to heterogeneous architectures. 
\item Design of a light weight algorithm for data shuffling required prior to performing parallel FFT in distributed memory systems.
\end{itemize}
Data shuffling is required due to the chunked wise storage of the data. Before performing FFT, it is required to shuffle the data such that the arrangement is suitable for FFT transform. It is beneficial to perform this operations with maximum parallelism and minimum extra container usage. 
\section{Mathematical Formulation}
Poisson's equation given in Eq. \ref{eq:poisson} in three-dimensional space is discretized using a second-order accurate central finite difference scheme on a directionally-uniform cell-centered grid,
\begin{equation}
\begin{aligned}
\frac{u_{i+1,j,k}-2 u_{i,j,k}+u_{i-1,j,k}}{\Delta x^{2}}+ 
\frac{u_{i,j+1,k}-2 u_{i,j,k}+u_{i,j-1,k}}{\Delta y^{2}}+ \\
\frac{u_{i,j,k+1}-2 u_{i,j,k}+u_{i,j,k-1}}{\Delta z^{2}}=f_{i,j,k}
\end{aligned}
\end{equation}
where $u_{i}$ is the independent variable and $f$ is the source term.
After incorporating the boundary conditions, the discretization will lead to the system of algebraic equation 
\begin{equation}
Ax=b
\end{equation}
This operator has a bandwidth of 7. However, under certain boundary conditions, the Laplacian operator may be decomposed into the eigenvectors using appropriate Fourier transforms \cite{schumann1988fast}. Therefore, this approximation in a sense, is similar to the pseudo-spectral approximation but uses different set of eigenvalues. The efficiency of the algorithm comes from the efficiency of FFT as well as
its potential for fine-grain parallelism.
\begin{figure}[h!]
\centering
\begin{tikzpicture}[scale=1.0][line cap=round,line join=round,>=triangle 45,x=1cm,y=1cm]
\end{tikzpicture}
\end{figure}

\subsection{Computational Complexity Analysis}
The first step in adopting any algorithm in heterogeneous cluster computing is to perform a cost analysis and investigate its amenability for different levels of parallelism. 
Most computational complexity analyses have been traditionally calculated for serial execution models. With ever increasing computing power, it is beneficial to revisit the cost analysis by considering parallelism. 
we quantify the algorithmic complexity of the FFT-based Poisson solver and demonstrate its potential for parallelism. 
Consider a grid with $N_{x}$,$N_{y}$ and $N_{z}$ points in each direction. 
Since the cost of an FFT for a given signal of length $N_{x}$ is proportional to $N_{x}log(N_{x})$ \cite{cooley1970fast}.
The cost of two consecutive FFT transformations in $x$ and $y$ directions is given as,

\begin{equation*}
\begin{aligned}
 & N_{y}N_{z}(\mathcal{O} (N_{x}log(N_{x})) + \\
       & N_{x}N_{z}(\mathcal{O} (N_{y} log(N_{y}))
\end{aligned}
\end{equation*}
These transformations yield  $\mathcal{O} (N_{x} N_{y})$ tridiagonal systems that can be solved using the Thomas algorithm which has complexity of 
\begin{equation*}
N_{x}N_{y}(\mathcal{O} (N_{z})) 
\end{equation*}
Complexity of inverse transforms is the same as the direct transforms, therefore the total cost is
\begin{equation*}
\begin{aligned}
 & 2 N_{y}N_{z}(\mathcal{O} (N_{x}log(N_{x})) + \\
       & 2 N_{x}N_{z}(\mathcal{O} (N_{y} log(N_{y})) + \\ 
       & N_{x}N_{y}(\mathcal{O} (N_{z})). 
\end{aligned}
\end{equation*}
For the sake of simplicity, assume  $N_{x}=N_{y}=N_{z}=N$, also ignoring the constant and the smaller order terms gives,
\begin{equation*}
N^{3}(\mathcal{O} (log(N))
\end{equation*}
However, FFT-based algorithm has a high potential for fine-grain parallelism. In fact, all the operations at every stage of the solution can be performed simultaneously in batches. 
We have reported the speed-up gained with a single GPU relative to a single CPU in \cite{JHasbestan2018}. 
 

\section{Boundary Conditions}
\begin{figure}[h!]
\centering
\subfloat[ DST-I]{
\begin{tikzpicture}[scale=0.4]
\begin{axis}[
width = 20cm,
height = 5cm,
hide x axis,
hide y axis,
]
\addplot + [blue,ycomb,mark options={fill=blue}] plot coordinates
{
(0.03978,1.00000)
(1.34313,0.92334)
(2.44661,0.83822)
(3.58791,0.73668)
(4.73067,0.64749)
(5.99867,0.60783)
(7.12523,0.54948)
(8.19310,-0.00070)
(9.27841,-0.57144)
(10.54469,-0.62551)
(11.57451,-0.66133)
(12.79634,-0.75455)
(14.01719,-0.85599)
(15.08064,-0.94322)
(16.30542,-1.01174)
(17.47471,-1.04534)
(18.65357,-0.99869)
(19.88817,0.01511)
};
\addplot + [red,ycomb,mark options={fill=red}] plot coordinates
{
(21.12105,1.01450)
(22.32005,1.06323)
(23.46822,1.01932)
(24.69349,0.95491)
(25.79648,0.86568)
(26.93728,0.76002)
(28.06065,0.67492)
(29.28838,0.63110)
(30.45545,0.57897)
};
\addplot + [blue,ycomb,mark options={fill=blue}] plot coordinates
{(31.5,-0.00070)
(32.70808,-0.54184)
(33.89480,-0.59600)
(34.96415,-0.63383)
(36.14571,-0.73121)
(37.36755,-0.82442)
(38.46980,-0.91984)
(39.67444,-0.99044)
(40.84471,-1.01581)
(42.00344,-0.97124)
(43.23755,0.03845)
(44.47092,1.04195)
(45.62891,1.08035)
(46.77855,1.04878)
(48.00358,0.98232)
(49.10657,0.89308)
(50.20857,0.79561)
(51.43041,0.70240)
};
\end{axis}
\end{tikzpicture}}

\subfloat[ DST-II]{
\begin{tikzpicture}[scale=0.4]
\begin{axis}[
width = 20cm,
height = 5cm,
hide x axis,
hide y axis,
]
\addplot + [blue,ycomb,mark options={fill=blue}] plot coordinates
{
(0.03978,1.00000)
(1.34313,0.92334)
(2.44661,0.83822)
(3.58791,0.73668)
(4.73067,0.64749)
(5.99867,0.60783)
(7.12523,0.54948)
(8.27841,-0.57144)
(9.54469,-0.62551)
(10.57451,-0.66133)
(11.79634,-0.75455)
(13.01719,-0.85599)
(14.08064,-0.94322)
(15.30542,-1.01174)
(16.47471,-1.04534)
(17.65357,-0.99869)
};
\addplot + [red,ycomb,mark options={fill=red}] plot coordinates
{
(19.12105,1.01450)
(20.32005,1.06323)
(21.46822,1.01932)
(22.69349,0.95491)
(23.79648,0.86568)
(24.93728,0.76002)
(26.06065,0.67492)
(27.28838,0.63110)
(28.45545,0.57897)
};
\addplot + [blue,ycomb,mark options={fill=blue}] plot coordinates
{
(29.70808,-0.54184)
(30.89480,-0.59600)
(31.96415,-0.63383)
(33.14571,-0.73121)
(34.36755,-0.82442)
(35.46980,-0.91984)
(36.67444,-0.99044)
(37.84471,-1.01581)
(39.00344,-0.97124)

(40.47092,1.04195)
(41.62891,1.08035)
(42.72,1.04878)
(43.82,0.98232)
(44.92,0.89308)
(46,0.79561)
(46.95,0.70240)
};
\end{axis}
\end{tikzpicture}}

\subfloat[ DST-III]{
\begin{tikzpicture}[scale=0.4]
\begin{axis}[
width = 20cm,
height = 5cm,
hide x axis,
hide y axis,
]
\addplot + [blue,ycomb,mark options={fill=blue}] plot coordinates
{
(0.40937, -10.10479)
(1.77231, -9.52345)
(3.44192, -8.46789)
(4.64978, -7.46577)
(6.19225, -6.51569)
(8.02254, -5.98617)
(9.18918, -5.45781)
(10.95380, -6.19155)
(12.58010, -6.82028)
(14.08485, -7.65974)
(15.60078, -8.70971)
(17.09994, -9.44393)
(18.32806, -10.07335)
(19.81045, -10.49180)
(21.2, -9.85679)
(22.5, 0.06613)
};
\addplot + [red,ycomb,mark options={fill=red}] plot coordinates
{
(23.38301, 9.72499)
(24.95902, 10.04355)
(26.41626, 10.09873)
(28.05374, 9.25950)
(29.56967, 8.20953)
(31.14638, 7.26493)
(32.58197, 6.47797)
(34.06157, 6.11215)
};
\addplot + [blue,ycomb,mark options={fill=blue}] plot coordinates
{
(35.74865, 5.58879)
(36.99003, 5.95939)
(38.82032, 6.48892)
(40.10292, 7.33327)
(41.57903, 8.28323)
(43.04955, 9.33844)
(44.54522, 9.92001)
(46.11284, 10.39645)
(47.53446, 9.87262)

(49.0866, -0.19188)
(51, -9.90965)
(52.85488, -10.32857)
(54.63278, -10.06229)
(56.05649, -9.37558)
(57.52981, -8.37300)
(58.80402, -7.37076)
(60.21097, -6.36829)
};
\end{axis}
\end{tikzpicture}}

\subfloat[ DST-IV]{
\begin{tikzpicture}[scale=0.4]
\begin{axis}[
width = 20cm,
height = 5cm,
hide x axis,
hide y axis,
]
\addplot + [blue,ycomb,mark options={fill=blue}] plot coordinates
{
(0.13187, -6.95234)
(1.58242, -6.47978)
(2.96703, -5.89913)
(4.48352, -5.20991)
(6.00000, -4.62895)
(7.58242, -4.26431)
(8.96703, -3.86408)
(10.61538, -3.89619)
(12.00000, -4.25369)
(13.58242, -4.57462)
(15.09890, -5.18438)
(16.54945, -5.86646)
(18.00000, -6.54853)
(19.58242, -6.97772)
(20.96703, -7.19088)
(22.54945, -6.86233)
};
\addplot + [red,ycomb,mark options={fill=red}] plot coordinates
{
(23.86813, 6.74393)
(25.31868, 7.00000)
(26.83516, 6.78715)
(28.35165, 6.32173)
(29.93407, 5.71213)
(31.38462, 5.06614)
(32.83516, 4.45623)
(34.41758, 4.06312)
(35.80220, 3.77780)
};
\addplot + [blue,ycomb,mark options={fill=blue}] plot coordinates
{
(37.38462, 3.74552)
(38.83516, 4.07375)
(40.35165, 4.43822)
(41.93407, 5.05543)
(43.31868, 5.74433)
(44.96703, 6.36170)
(46.28571, 6.79786)
(47.93407, 7.12657)
(49.31868, 6.80515)
(50.90110, -6.79413)
(52.08791, -7.15210)
(53.60440, -6.85979)
(55.31868, -6.42268)
(56.70330, -5.84203)
(58.21978, -5.11673)
(59.67033, -4.42768)

};
\end{axis}
\end{tikzpicture}}
\caption{Different versions of Discrete Sine Transforms (DST).}
\label{fig:DST}
\end{figure}
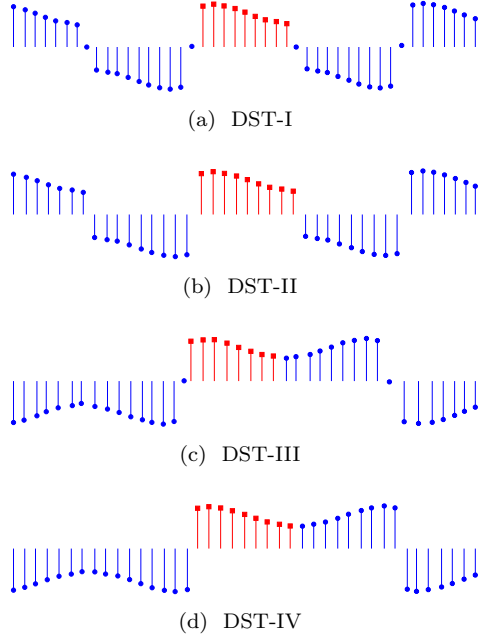

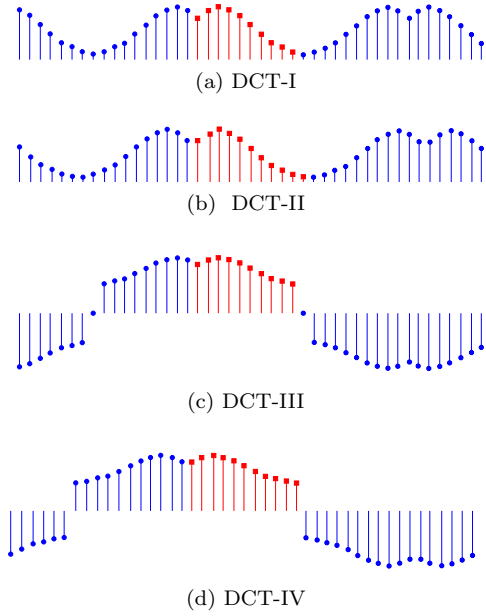
\begin{figure}
\centering
\subfloat[DCT-I]{
\begin{tikzpicture}[scale=0.4]
\begin{axis}[
width = 20cm,
height = 3.5cm,
hide x axis,
hide y axis,
]
\addplot + [blue,ycomb,mark options={fill=blue}] plot coordinates
{
(0.00000, 9.83594)
(1.31868, 9.26036)
(2.68446, 8.27489)
(3.95604, 7.28955)
(5.41601, 6.34493)
(6.78179, 5.93323)
(8.10047, 5.39864)
(9.51334, 5.15081)

(10.92622, 5.39478)
(12.29199, 5.92570)
(13.56358, 6.29282)
(14.88226, 7.27462)
(16.34223, 8.21525)
(17.61381, 9.19712)
(19.02669, 9.80994)
(20.39246, 10.13595)
(21.71115, 9.76529)

};
\addplot + [red,ycomb,mark options={fill=red}] plot coordinates
{

(22.98273, 8.94388)
(24.39560, 9.76163)
(25.66719, 10.16972)
(27.03297, 9.83999)
(28.44584, 9.22331)
(29.76452, 8.27888)
(31.22449, 7.25230)
(32.49608, 6.34892)
(33.90895, 5.89617)
(35.27473, 5.36152)
};
\addplot + [blue,ycomb,mark options={fill=blue}] plot coordinates
{

(36.64050, 5.07277)
(38.05338, 5.35772)
(39.46625, 5.76563)
(40.78493, 6.33760)
(42.15071, 7.23737)
(43.51648, 8.13714)
(44.88226, 9.11888)
(46.20094, 9.73184)
(47.56672, 10.09882)
(48.83830, 9.72823)
(50.34537, 8.94749)
(51.47567, 9.68365)
(52.84144, 10.13260)
(54.25432, 9.76182)
(55.52590, 9.10434)
(56.98587, 8.15973)
(58.35165, 7.21524)
(59.62323, 6.27088)
};
\end{axis}
\end{tikzpicture}}

\subfloat[ DCT-II]{
\begin{tikzpicture}[scale=0.4]
\begin{axis}[
width = 20cm,
height = 3.5cm,
hide x axis,
hide y axis,
]
\addplot + [blue,ycomb,mark options={fill=blue}] plot coordinates
{
(-0.09852, 6.86793)
(1.36132, 6.00622)
(2.68143, 5.34294)
(4.14754, 4.94349)
(5.42377, 4.54445)
(6.79762, 4.34331)
(8.17327, 4.27425)
(9.50593, 4.53547)
(10.98233, 4.89542)
(12.36469, 5.32163)
(13.75108, 6.04499)
(15.13882, 6.86741)
(16.52476, 7.55775)
(17.86100, 8.08311)
(19.14664, 8.37745)
(20.51915, 8.07726)
(21.74434, 7.41419)
};
\addplot + [red,ycomb,mark options={fill=red}] plot coordinates
{
(23.12088, 7.41116)
(24.74236, 7.96891)
(25.93486, 8.39553)
(27.25901, 8.02941)
(28.62883, 7.53111)
(30.08956, 6.73544)
(31.40789, 5.94008)
(32.72801, 5.27680)
(34.14664, 4.87745)
(35.51826, 4.51122)
(36.89211, 4.31009)
};
\addplot + [blue,ycomb,mark options={fill=blue}] plot coordinates
{

(38.22074, 4.27414)
(39.64743, 4.46912)
(40.98188, 4.86241)
(42.41037, 5.18945)
(43.84467, 5.94573)
(45.27943, 6.73502)
(46.52387, 7.49172)
(47.95415, 7.95083)
(49.38218, 8.24486)
(50.70723, 7.94477)
(51.98033, 7.31461)
(53.40389, 7.27846)
(54.69400, 7.90298)
(56.17040, 8.26293)
(57.44843, 7.99597)
(58.86438, 7.39851)
(60.08956, 6.73544)

};
\end{axis}
\end{tikzpicture}}

\subfloat[DCT-III]{
\begin{tikzpicture}[scale=0.4]
\begin{axis}[
width = 20cm,
height = 6cm,
hide x axis,
hide y axis,
]
\addplot + [blue,ycomb,mark options={fill=blue}] plot coordinates
{
(0.00000, -9.83594)
(1.31868, -9.26036)
(2.68446, -8.27489)
(3.95604, -7.28955)
(5.41601, -6.34493)
(6.78179, -5.93323)
(8.10047, -5.39864)
(9.51334, -0)

(10.92622, 5.39478)
(12.29199, 5.92570)
(13.56358, 6.29282)
(14.88226, 7.27462)
(16.34223, 8.21525)
(17.61381, 9.19712)
(19.02669, 9.80994)
(20.39246, 10.13595)
(21.71115, 9.76529)

};
\addplot + [red,ycomb,mark options={fill=red}] plot coordinates
{

(22.98273, 8.94388)
(24.39560, 9.76163)
(25.66719, 10.16972)
(27.03297, 9.83999)
(28.44584, 9.22331)
(29.76452, 8.27888)
(31.22449, 7.25230)
(32.49608, 6.34892)
(33.90895, 5.89617)
(35.27473, 5.36152)
};
\addplot + [blue,ycomb,mark options={fill=blue}] plot coordinates
{

(36.64050, 0)
(38.05338, -5.35772)
(39.46625, -5.76563)
(40.78493, -6.33760)
(42.15071, -7.23737)
(43.51648, -8.13714)
(44.88226, -9.11888)
(46.20094, -9.73184)
(47.56672, -10.09882)
(48.83830, -9.72823)
(50.34537, -8.94749)
(51.47567, -9.68365)
(52.84144, -10.13260)
(54.25432, -9.76182)
(55.52590, -9.10434)
(56.98587, -8.15973)
(58.35165, -7.21524)
(59.62323, -6.27088)
};
\end{axis}
\end{tikzpicture}}

\subfloat[DCT-IV]
{
\begin{tikzpicture}[scale=0.4]
\begin{axis}[
width = 20cm,
height = 6cm,
hide x axis,
hide y axis,
]
\addplot + [blue,ycomb,mark options={fill=blue}] plot coordinates
{
(-0.24344, -5.53397)
(1.18905, -4.88882)
(2.66946, -4.21558)
(4.04412, -3.96542)
(5.37543, -3.57402)
(6.70065, -3.40841)
(8.20997, 3.58733)
(9.63028, 3.78091)
(11.05744, 4.22849)
(12.38418, 4.45055)
(13.81514, 5.03925)
(15.34424, 5.76881)
(16.68163, 6.38600)
(17.92012, 6.83412)
(19.24838, 7.11263)
(20.75010, 6.82613)
(22.00913, 6.28629)
};
\addplot + [red,ycomb,mark options={fill=red}] plot coordinates
{
(23.28186, 6.25447)
(24.57056, 6.81535)
(26.13465, 7.09319)
(27.35337, 6.80749)
(28.75770, 6.40837)
(30.06162, 5.78372)
(31.55116, 5.04564)
(32.85508, 4.42100)
(34.16812, 4.13503)
(35.57474, 3.82058)
(36.88855, 3.56284)
};
\addplot + [blue,ycomb,mark options={fill=blue}] plot coordinates
{
(38.11487, -3.44062)
(39.56942, -3.72699)
(40.83378, -4.06927)
(42.14530, -4.41168)
(43.59300, -4.95205)
(44.84595, -5.71769)
(46.24496, -6.31438)
(47.50780, -6.71310)
(48.91366, -7.05578)
(50.29137, -6.69273)
(51.62571, -6.18844)
(53.04070, -6.19243)
(54.20616, -6.70378)
(55.75428, -7.01864)
(57.08178, -6.76835)
(58.41537, -6.29228)
(59.75200, -5.70331)
};
\end{axis}
\end{tikzpicture}
}
\caption{Different versions of Discrete Cosine Transform (DCT).}
\label{fig:DCT}
\end{figure}

Type of boundary conditions imposed on the Poisson's equation affects the details of the overall algorithm and therefore deserves a separate discussion. Application of the $\mathcal{L}$ to the Fourier basis function leads to
\begin{equation}
\mathcal{L} q =\lambda q 
\end{equation}
where $\lambda$ is the eigenvalue and $q$ is the eigenfunction. Therefore, the linear operator can also be regarded as an eigenfunction decomposition. 
In eigenfunction decomposition, due to the linearity of the operator, each boundary condition need to be satisfied by each eigenfunction.
Therefore depending on the boundary condition a different transformation is selected.
For Dirichlet and Neumann boundary conditions, cosine and sine transforms are used, respectively. 
These transforms in the real domain may be obtained from complex transforms by properly extending or modifying the signal.
There are several possibilities to perform this extension at the boundaries and each of these extensions leads to different variants of real-to-real transforms \cite{martucci1994symmetric}. There are eight possibilities for Discrete Sine Transforms (DST) and Discrete Cosine Transforms (DCT).
Figure \ref{fig:DST} and \ref{fig:DCT} provide four most popular real-to-real transformations for DST and DCT, respectively. The red represents the original signal while the blue demonstrates the extended signal needed to perform each transformation. Figures \ref{fig:DST} and \ref{fig:DCT} are presented for illustration only and they do not represent the algorithms used in \texttt{PittPack}. No signal extensions are performed in \texttt{PittPack}.

The selection of the transformations are done depending on the boundary conditions of the problem under consideration and will be elaborated in the upcoming sections.

\texttt{PittPack} performs complex FFT transform for all the transformations.
The reason to do this is two fold. First, \texttt{cuFFT} library does not support real-to-real
transform, mainly due to the fact that these transformations can all be realized using complex transforms
at a minor computational and memory cost. Second, it enables more elegant software design.
The procedure for calculating DST and DCT using complex Fourier transforms can be divided in three steps:

\begin{enumerate}
\item Pre-processing : Modify the original input signal
\item Complex Fourier transformation : Perform FFT on the modified signal
\item Post-processing: Extract the desired real transformation 
\end{enumerate}
where different types of the real transforms require different operations at every stage.
In the following sections we elaborate on the details of these transformations 
used in the \texttt{PittPack} software. 

\definecolor{ffvvqq}{rgb}{1,0.3333333333333333,0}
\definecolor{ududff}{rgb}{0.30196078431372547,0.30196078431372547,1}
\begin{figure}[h!]
\centering
\begin{tikzpicture}[scale=1.0][line cap=round,line join=round,>=triangle 45,x=1cm,y=1cm]
\clip(-6.57846340992715,-0.430297704863707) rectangle (1.816478237548978,4.413901575962472);
\draw [line width=2pt] (-5,4)-- (0,4);
\draw [line width=2pt] (-5,4)-- (-5,0);
\draw [line width=2pt] (0,4)-- (0,0);
\draw [line width=2pt] (-5,0)-- (0,0);
\draw [line width=2pt] (-4,3)-- (-3,3);
\draw [line width=2pt] (-2,3)-- (-1,3);
\draw [line width=2pt] (0,3)-- (-1,3);
\draw [line width=2pt] (0,2)-- (-1,2);
\draw [line width=2pt] (-1,2)-- (-2,2);
\draw [line width=2pt] (-2,2)-- (-3,2);
\draw [line width=2pt] (-3,2)-- (-4,2);
\draw [line width=2pt] (-4,2)-- (-5,2);
\draw [line width=2pt] (-4,3)-- (-5,3);
\draw [line width=2pt] (-3,3)-- (-2,3);
\draw [line width=2pt] (-2,3)-- (-2,4);
\draw [line width=2pt] (-1,3)-- (-1,4);
\draw [line width=2pt] (-3,3)-- (-3,4);
\draw [line width=2pt] (-4,3)-- (-4,4);
\draw [line width=2pt] (-4,2)-- (-4,3);
\draw [line width=2pt] (-3,2)-- (-3,3);
\draw [line width=2pt] (-2,2)-- (-2,3);
\draw [line width=2pt] (-1,2)-- (-1,3);
\draw [line width=2pt] (0,1)-- (-1,1);
\draw [line width=2pt] (-1,1)-- (-1,2);
\draw [line width=2pt] (-2,2)-- (-2,1);
\draw [line width=2pt] (-2,1)-- (-1,1);
\draw [line width=2pt] (-1,0)-- (-1,1);
\draw [line width=2pt] (-2,1)-- (-2,0);\draw [line width=2pt] (-3,0)-- (-3,1);\draw [line width=2pt] (-3,1)-- (-2,1);\draw [line width=2pt] (-3,2)-- (-3,1);\draw [line width=2pt] (-3,1)-- (-4,1);\draw [line width=2pt] (-4,1)-- (-4,2);\draw [line width=2pt] (-4,1)-- (-4,0);\draw [line width=2pt] (-4,1)-- (-5,1);\begin{scriptsize}\draw [fill=ududff] (-4.5,0.5) circle (2.5pt);\draw [fill=ududff] (-3.5,0.5) circle (2.5pt);\draw [fill=ududff] (-2.508737791134487,0.47997746055597285) circle (2.5pt);\draw [fill=ududff] (-1.5,0.5) circle (2.5pt);\draw [fill=ududff] (-0.48018782870022714,0.47997746055597285) circle (2.5pt);\draw [fill=ududff] (-0.5,1.5) circle (2.5pt);\draw [fill=ududff] (-0.5,2.5) circle (2.5pt);\draw [fill=ududff] (-1.4719233658903097,2.5085274229902326) circle (2.5pt);\draw [fill=ududff] (-1.5,1.5) circle (2.5pt);\draw [fill=ududff] (-2.5,1.5) circle (2.5pt);\draw [fill=ududff] (-2.5,2.5) circle (2.5pt);\draw [fill=ududff] (-3.5,2.5) circle (2.5pt);\draw [fill=ududff] (-3.5,1.5) circle (2.5pt);\draw [fill=ududff] (-4.5,1.5) circle (2.5pt);\draw [fill=ududff] (-4.5,2.5) circle (2.5pt);\draw [fill=ududff] (-4.5,3.5) circle (2.5pt);\draw [fill=ududff] (-2.5,3.5) circle (2.5pt);\draw [fill=ududff] (-1.5019759579263727,3.500262960180315) circle (2.5pt);\draw [fill=ududff] (-0.5,3.5) circle (2.5pt);\draw [fill=ududff] (-1.5019759579263727,3.500262960180315) circle (2.5pt);\draw [fill=ududff] (-3.5,3.5) circle (2.5pt);\draw [color=ffvvqq] (-5,0.5042640427990959)-- ++(-4.5pt,-4.5pt) -- ++(9pt,9pt) ++(-9pt,0) -- ++(9pt,-9pt);\draw [color=ffvvqq] (-5,1.5054788590730652)-- ++(-4.5pt,-4.5pt) -- ++(9pt,9pt) ++(-9pt,0) -- ++(9pt,-9pt);\draw [color=ffvvqq] (-5,2.5066936753470346)-- ++(-4.5pt,-4.5pt) -- ++(9pt,9pt) ++(-9pt,0) -- ++(9pt,-9pt);\draw [color=ffvvqq] (-5,3.522632238919151)-- ++(-4.5pt,-4.5pt) -- ++(9pt,9pt) ++(-9pt,0) -- ++(9pt,-9pt);\draw [color=ffvvqq] (0,3.5)-- ++(-4.5pt,-4.5pt) -- ++(9pt,9pt) ++(-9pt,0) -- ++(9pt,-9pt);\draw [color=ffvvqq] (0,2.502930304321515)-- ++(-4.5pt,-4.5pt) -- ++(9pt,9pt) ++(-9pt,0) -- ++(9pt,-9pt);\draw [color=ffvvqq] (0,1.5009914945503509)-- ++(-4.5pt,-4.5pt) -- ++(9pt,9pt) ++(-9pt,0) -- ++(9pt,-9pt);\draw [color=ffvvqq] (0,0.48102678425537304)-- ++(-4.5pt,-4.5pt) -- ++(9pt,9pt) ++(-9pt,0) -- ++(9pt,-9pt);\draw [color=ffvvqq] (-4.5,4)-- ++(-4.5pt,-4.5pt) -- ++(9pt,9pt) ++(-9pt,0) -- ++(9pt,-9pt);\draw [color=ffvvqq] (-3.5,4)-- ++(-4.5pt,-4.5pt) -- ++(9pt,9pt) ++(-9pt,0) -- ++(9pt,-9pt);\draw [color=ffvvqq] (-2.5,4)-- ++(-4.5pt,-4.5pt) -- ++(9pt,9pt) ++(-9pt,0) -- ++(9pt,-9pt);\draw [color=ffvvqq] (-1.5,4)-- ++(-4.5pt,-4.5pt) -- ++(9pt,9pt) ++(-9pt,0) -- ++(9pt,-9pt);\draw [color=ffvvqq] (-0.5,4)-- ++(-4.5pt,-4.5pt) -- ++(9pt,9pt) ++(-9pt,0) -- ++(9pt,-9pt);\draw [color=ffvvqq] (-4.5,0)-- ++(-4.5pt,-4.5pt) -- ++(9pt,9pt) ++(-9pt,0) -- ++(9pt,-9pt);\draw [color=ffvvqq] (-3.5,0)-- ++(-4.5pt,-4.5pt) -- ++(9pt,9pt) ++(-9pt,0) -- ++(9pt,-9pt);\draw [color=ffvvqq] (-2.5,0)-- ++(-4.5pt,-4.5pt) -- ++(9pt,9pt) ++(-9pt,0) -- ++(9pt,-9pt);
\draw [color=ffvvqq] (-1.5,0)-- ++(-4.5pt,-4.5pt) -- ++(9pt,9pt) ++(-9pt,0) -- ++(9pt,-9pt);\draw [color=ffvvqq] (-0.5,0)-- ++(-4.5pt,-4.5pt) -- ++(9pt,9pt) ++(-9pt,0) -- ++(9pt,-9pt);
\end{scriptsize}
\end{tikzpicture}
\caption{Illustration of the cell-centered arrangement of the solution variable and the corresponding boundary faces}
\label{fig:boundary_ill}
\end{figure}
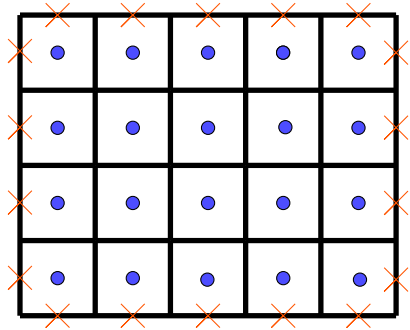
\noindent

\subsection{Periodic Boundary Condition}
Discrete Fourier Transform is calculated as,
\begin{equation}
Y_k=\sum_{n=0}^{N-1} u_n e^{- i(\frac{2\pi}{N})( k) (n)}
\end{equation}
\noindent
where $i= \sqrt{(-1)}$. Periodic boundary condition are naturally satisfied by the basis functions of the complex Fourier transform. No additional pre- or post- processing stages are required for this type of boundary conditions. 

\subsection{Neumann-type Boundary Condition}
Cosine transformation is used in this situation. Since Neumann-type boundary condition should be 
satisfied at the cell-face, Type II discrete cosine transformation (DCT10) is used for this case.
The most memory efficient method is introduced by Makhoul \cite{Makhoul}. 
He uses Hermitian symmetry property to construct the transformation.
DCT10 is mathematically expressed as,
\begin{equation}
Y_k=\sum_{n=0}^{N-1} u_n cos\big[{(n+\frac{1}{2})}(k)(\frac{\pi}{N})\big]
\end{equation}
For sake of completeness, we give a brief overview of the algorithm;
\newline
Given a signal $\mathbf{u}=[u_1, u_2, u_3, u_4, u_5]$ with size $N$, following steps will calculate DCT10;

\begin{itemize}
\item Construct $y=[u_1, u_3, u_5, u_4, u_2 ]$, in other words, take the even points in order then append the odd points in reverse order  
\item Perform complex FFT on $y$ 
\item Enforce Hermitian symmetry
\item Scale the result by $2 e^{ (-k \pi \sqrt{-1} )/(2N)}$  
\item Retain and use the real part of the transformation
\end{itemize}

\noindent
The inverse transformation of Type I (DCT10) is obtained by performing Type II transformation, i.e. DCT01.
DCT01 is mathematically defined as,
\begin{equation}
Y_k=\frac{1}{2}u_0+\sum_{n=1}^{N-1} u_n cos\big[{(k+\frac{1}{2})}(n)\frac{ \pi }{N}\big]
\end{equation}
The procedure for DCT01 is summarized as follows,
\begin{itemize}
\item Calculate  $V=\frac{1}{2} e^{(-k \pi \sqrt{-1} )/(2 N)}(u[k]-u[N-k])$
\item Perform complex inverse FFT (IFFT) on $V$ 
\item Retrieve the data assuming that it has been previously rearranged by a DCT10 transform 
\end{itemize}

\subsection{Dirichlet-type Boundary Condition}
In this case, the non-homogeneous boundary values are prescribed at the cell faces.
We first employ the boundary conditions to the finite difference equations.
By assuming linear extrapolation, the value at the neighboring ghost cell can easily be calculated
and moved to the right hand side. This essentially updates the source term and constructs a new signal.
We then employ Discrete Sine Transform Type II (DST10), which is an odd extension at both edges of the signal
with respect to the cell-face. This signal is related to cosine transform DCT10. With the infrastructure already in place to perform DCT10, calculating DST10 is straightforward. 
DST10 is defined as,
\begin{equation}
Y_k=\sum_{n=0}^{N-1} u_n sin\big[{(n+\frac{1}{2})}(k+1)\frac{\pi }{N}\big]
\end{equation}
Here, we briefly explain the procedure.
Given a signal with $u=[u_1,u_2, u_3, u_4, u_5 ]$ entries,

\begin{itemize}
\item Construct $u=[u_5, -u_4, u_3, -u_2, u_1 ]$
\item Apply DCT10 on $u$ to obtain $U$
\item Rearrange $U$ to get $DST10(u)=[U_5, -U_4, U_3, -U_2, U_1]$
\end{itemize}
\noindent
The inverse of the DST10, is DCT01 and is mathematically given as
\begin{equation}
Y_k=\frac{(-1)^{k}}{2}u_n-1+\sum_{k=0}^{N-2} u_n cos\big[(n+1) (k+\frac{1}{2})\frac{ \pi}{N}\big]
\end{equation}
Similarly, to perform the inverse transform, the relation between the DST01 and DCT01 is used.
The procedure for this is very similar to the forward transform and is presented as follows,
\begin{itemize}
\item Construct $u=[u_5, -u_4, u_3, -u_2, u_1]$
\item Perform DCT01 $u$ to obtain $U$
\item Rearrange $U$ to get $DST01(u)=[U_5, -U_4, U_3, -U_2, U_1]$
\end{itemize}
\noindent
Note that same method is used to perform the first and last steps.
Meaning that rearranging the same array twice will recover the original signal.

\noindent
The summary of all the transformations and their inverse are listed in Table \ref{tab:TransformationList}
\begin{table}[h!]
\begin{center}
 \begin{tabular}{c|c|c} 
      & Forward & Backward  \\ [0.7ex] 
 \hline
  Dirichlet  &  DST10 & DST01 \\ [0.7ex]
 \hline 
 Neumann     &  DCT10 & DCT01\\ [0.7ex]
 \hline
 Periodic   &  FFT & IFFT\\ [0.7ex]
  \end{tabular}
 \caption{Summary of the appropriate Fourier transformations based on the boundary conditions}
 \label{tab:TransformationList} 
\end{center}
\end{table}

\subsection{Eigenvalues}
Eigenvalues for each transformation at a given direction is listed in Table \ref{tab:eig}.
To obtain the eigenvalue in two directions we simply sum up the eigenvalues in each direction.
\begin{table}[h!]
\begin{center}
 \begin{tabular}{c|c} 
 Boundary  Condition & Eigenvalue\\ [0.7ex] 
 \hline
 Periodic & $-2+2$ $cos(2k \pi/n)$  \\ [0.7ex] 
 \hline
 Neumann-Neumann &  -$2+2$ $cos(k \pi/n)$ \\ [0.7ex] 
 \hline
 Dirichlet-Dirichlet & $-2+2$ $cos((k+1) \pi/n)$ \\ [0.7ex] 
 \end{tabular}
 \caption{Eigenvalues of the discrete system based on the boundary conditions}
 \label{tab:eig}
\end{center}
\end{table}

\section{Solution of the Resulting Tridiagonal Matrices}
After performing two consecutive transformations in $x$ and $y$ directions, discrete Poisson's equation reduces to a system of tridiagonal equations in the $z$-direction, each with a different diagonal value. It is preferred to solve these equations directly. This is mainly due to two reasons. First, direct methods do not depend on the condition number as the procedure is performed only once. Second, for the system obtained after two Fourier transformations the change in the diagonal elements in refined meshes can act as destabilizing perturbation which in turn can deteriorate or even break the iterative solution strategies such as multigrid.       
As a result of the pencil decomposition, every processor owns a tridiagonal system of size $nxChunk \times nyChunk$,  which is local to each GPU (or CPU) and hence, no communication is needed at this stage of the solution. \texttt{PittPack} implements the most widely used direct solve algorithms under a class called \texttt{TriDiag}. The user can select the desired algorithm for the solution of the tridiagonal system. 
Giles et al. \cite{giles2014gpu,laszlo2014methods} give a concise review of the computational cost for various tridiagonal solvers on GPUs. Chang \cite{chang2014guide} et al.  report a guideline for implementing tridiagonal solvers on GPUs with CUDA.
Here we briefly review the most common algorithms and point out their advantage and disadvantages as applicable to extreme scale computing.    

\subsection{Thomas Algorithm}
Thomas algorithm is the most common method to solve a tridiagonal matrix. The method is presented in Algorithm \ref{alg:Thomas} for completeness. 
Fortunately, applying two FFT transforms on the finite difference stencil for the Poisson's equation leads to the 
tridiagonal system that are Toeplitz matrices. Therefore, we can specialize the general Thomas's algorithm to take advantage of this property and enhance its performance and reduce the memory consumption significantly.
In this approach, we store only 3 elements for each of the subdiagonal, superdiagonal and diagonal entries, as all the elements between the first and last element are the same. Also, the boundary conditions in the $z$-direction affect only the first and last entries of the tridiagonal system, except when periodic boundary conditions are applied in the $z$-direction. However, for that case, one may adopt the Sherman-Morrison modification \cite{sherman1950adjustment} of the coefficient matrix and break down the periodic boundary condition to two solves. Therefore, without loss of generality we can pursue the Thomas algorithm for our application. Detailed implementation of this method is given in  as given in Algorithm \ref{alg:Thomas_LowMem}.

\begin{algorithm}[h]
	\caption{Thomas Algorithm}
	\label{alg:Thomas}
	\begin{algorithmic}
	    \State {define $N$ : problem size}
		\State {define $a[N]$ : lower diagonal elements}
		\State {define $b[N]$ : diagonal elements}
		\State {define $c[N]$ : upper diagonal elements}
		\State {define $r[N]$ : right hand side}
		\State {define $\gamma[N]$ : auxiliary container}
		\State {define $beta$ : auxiliary variable}
		\State{$aux=b_{0}$}
		\State{$r_{0}=r_{0}/aux$}
		\State{$\gamma_{1}=c_{0}/b_{0}$}
		\State{\color{blue}{\textbf{Forward Pass}}}
		\For{ $ i=1, ..., N-1$}		
		\State{$\gamma_{i} = c_{i-1}/bet$}
		\State{$aux=b[i]-a_{i} \gamma_{i}$}
		\State{$r= r (r_{i}-a_{i} r_{i-1})/aux$}
		\EndFor		
		\State{\color{blue}{\textbf{Backward Pass}}}
		\For{ $i=N-2, ... , 0$}		
		\State{$r_{i}= - r_{i} + (\gamma_{i+1} r_{i+1})$}
		\EndFor		
		\State{Return $r$}
	\end{algorithmic}
\end{algorithm}

\begin{algorithm}[h]
	\caption{Thomas Algorithm for Toeplitz Matrices}
	\label{alg:Thomas_LowMem}
	\begin{algorithmic}
	    \State {define $N$ : problem size}
		\State {define $a[3]$ : lower diagonal elements}
		\State {define $b[3]$ : diagonal elements}
		\State {define $c[3]$ : upper diagonal elements}
		\State {define $r[N]$ : right hand side}
		\State {define $\gamma[N]$ : auxiliary container}
		\State {define $aux$ : auxiliary variable}
		\State{$aux=b_{0}$}
		\State{$r_{0}=r_{0}/b_{0}$}
		\State{$\gamma_{1}=c_{0}/b_{0}$}
		\State{\color{blue}{\textbf{Forward Pass}}}
		\For{ $ i=1, ..., N-1$}		
		\State{$\gamma_{i} = c_{1}/be$}
		\State{$aux=b_{1}-a_{1}*\gamma_{i}$}
		\State{$r= r (r_{i}-a_{1} r_{i-1})/aux$}
		\EndFor		
		\State{\color{blue}{\textbf{Backward Pass}}}
		\For{ $i=N-2, ... , 0$}		
		\State{$r_{i}= - r_{i} + (\gamma_{i+1} r_{i+1})$}
		\EndFor		
		\State{Return $r$}
	\end{algorithmic}
\end{algorithm}

Thomas algorithm is inherently serial because of the implicit data dependency. The implication of this aspect of Thomas algorithm for GPU computing is that only a single thread can be active for each tridiagonal system solution, preventing thread level parallelism per system. Inherently serial algorithms are not suitable to port to GPUs due to the limited computational power of a single thread. However, the good news for the current application is that, with we have many tridiagonal matrices. Therefore, it is possible to apply task parallelism at the gang (block) level. This yields significant gain in performance at a moderate increase in memory consumption. This approach is referred to as \textit{batch-solution mode} in \texttt{PittPack}. The user may apply one-dimensional as well as two-dimensional batches. A one-dimensional batch deploys number of blocks equal to the number of points in $nx$ or $ny$ direction. 
While a two-dimensional batch deploys blocks that are of size $nx \times ny$.
The first option is well suited for low-memory consumption while the latter provides maximum speed. The default setting chooses the two-dimensional batch option.


\subsection{Cyclic Reduction Technique}
This method is a divide and conquer type of algorithm \cite{sweet1974generalized}. At each level, it reduces the size of the system by eliminating the dependency of even-odd indexed equations. Cyclic reduction (CR) has the $\mathcal{O} (N_{z}log(N_{z}))$ complexity in serial. The cost of the CR algorithm when solved in parallel is $\mathcal{O}(log(N_{z}))$ \cite{bondeli1991divide}. We refer to this algorithm as Cyclic Reduction-Parallelized (CR-P) throughout this paper to emphasize the parallel implementation and distinguish it from the Parallel Cyclic Reduction (PCR) method \cite{zhang2010fast}, which is a different method that we explain in the next section.  

\begin{figure}[h]
    \centering
\begin{tikzpicture}[level distance=15mm]
  \tikzstyle{every node}=[fill=yellow,circle,inner sep=4pt]
  \tikzstyle{level 1}=[sibling distance=40mm,
    set style={{every node}+=[fill=green!50]}]
  \tikzstyle{level 2}=[sibling distance=40mm,
    set style={{every node}+=[fill=red!60!white]}]
  \tikzstyle{level 3}=[sibling distance=20mm,
    set style={{every node}+=[fill=blue!30]}]
  \node {3}[grow'=up]
     child {node {2}
       child {node {1}
         child {node {0}}
         child {node {1}}
         child {node {2}}
       }
       child {node {5}
         child {node {2}}
         child {node {3}}
         child {node {4}}
       }
     }
     child {node {4}
       child {node {3}
         child {node {2}}
         child {node {3}}
         child {node {4}}
       }
        child {node {5}
         child {node {4}}
         child {node {5}}
         child {node {6}}
       }
     };
\end{tikzpicture}
  \caption{Forward pass stage of the CR-P algorithm for a tridiagonal system with 7 unknown. Elements 0 and 6 are boundary values}
  \label{fig:ForwardPhase}
\end{figure}
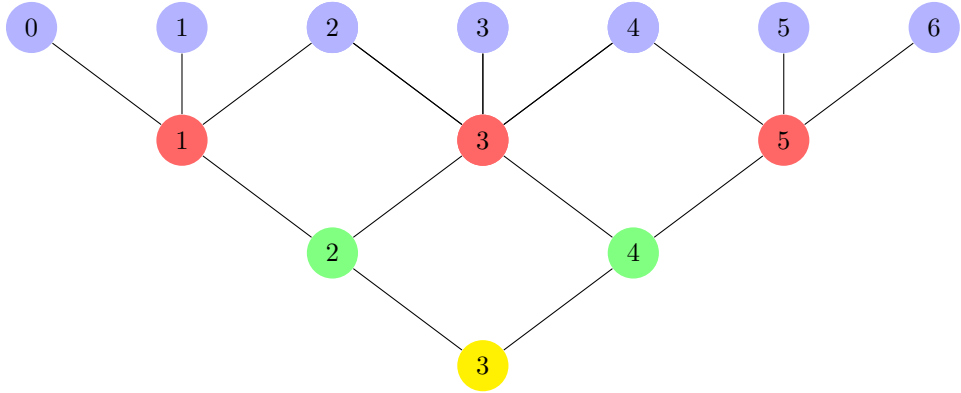

In CR, the tridiagonal system is normalized such that the diagonal elements are 1.0. The normalization prevents overflow as in diagonally dominant systems the value at the diagonal is bigger than one and successive multiplication could otherwise lead to overflow.
At every level, each of these operations can be performed in parallel. Figure \ref{fig:ForwardPhase} shows the forward pass of the algorithm for a system with seven equations. At each level the number of unknowns is reduced to 3, 2, and 1, respectively. 

\begin{figure}[h]
    \centering
\begin{tikzpicture}[level distance=15mm]
\filldraw[draw=none,fill=red!80!white ] ( -7, 4) circle[radius=0.3] node{$1$}; 
\filldraw[draw=none,fill=red!80!white ] ( -3, 4) circle[radius=0.3] node{$3$};
\filldraw[draw=none,fill=red!80!white ] ( 1, 4) circle[radius=0.3] node{$5$};

\filldraw[draw=none,fill=green!80!white ] ( -5, 2) circle[radius=0.3] node{$2$}; 
\filldraw[draw=none,fill=green!80!white ] ( -1, 2) circle[radius=0.3] node{$4$};
\draw[thick,->] (-.84,2.16) -- (0.82,3.82);
\draw[thick,->] (-4.84,2.16) -- (-3.16,3.84);
\draw[dashed] (-5.18,2.18) -- (-6.82,3.82);
\draw[dashed] (-1.16,2.16) -- (-2.84,3.84);
\end{tikzpicture}
  \caption{First phase of the backward substitution for CR-P algorithm}
  \label{fig:phase1backsub}
\end{figure}
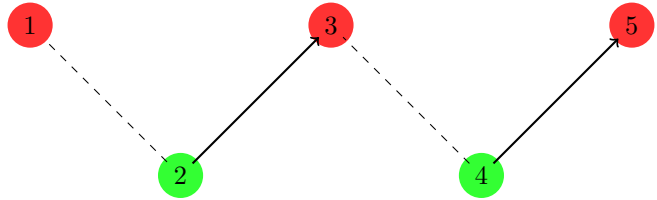

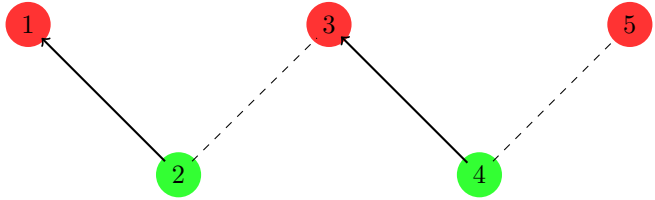
\begin{figure}[h]
    \centering
\begin{tikzpicture}[level distance=15mm]
\filldraw[draw=none,fill=red!80!white ] ( -7, 4) circle[radius=0.3] node{$1$}; 
\filldraw[draw=none,fill=red!80!white ] ( -3, 4) circle[radius=0.3] node{$3$};
\filldraw[draw=none,fill=red!80!white ] ( 1, 4) circle[radius=0.3] node{$5$};

\filldraw[draw=none,fill=green!80!white ] ( -5, 2) circle[radius=0.3] node{$2$}; 
\filldraw[draw=none,fill=green!80!white ] ( -1, 2) circle[radius=0.3] node{$4$};
\draw[thick,->] (-5.18,2.18) -- (-6.82,3.82);
\draw[thick,->] (-1.16,2.16) -- (-2.84,3.84);
\draw[dashed] (-.82,2.18) -- (0.82,3.82);
\draw[dashed] (-4.82,2.18) -- (-3.16,3.82);
\end{tikzpicture}
  \caption{Second phase of the backward substitution for CR-P algorithm}
  \label{fig:phase2backsub}
\end{figure}

\noindent
Once we reach the end of the forward pass, there is only one equation to solve, which is a trivial task. 
In the backward pass, this value is propagated to the previous levels.
The backward substitution pass at first glance seems data dependent.
However, a more detailed analysis reveals that the serial task may be split into two parallel sub-tasks as shown in Figures. \ref{fig:phase1backsub} and \ref{fig:phase2backsub}.

\subsection{Parallel Cyclic Reduction}
This algorithm eliminates the backward pass at the expense of extra memory utilization \cite{zhang2010fast}. 
The equations are normalized such that the diagonal elements are 1.0. The old values for the lower diagonal matrix and upper-diagonal matrices should not be overwritten. This leads to extra memory consumption compared to regular Thomas algorithm.  
One way to alleviate this problem is to use register memory per thread. This practice not only increases the occupancy on GPU, but also eliminates the gather operations from the global memory. The only draw back is that the maximum number of grid points for each system has to be less than 4096 for NVIDIA Tesla V100. This is mainly due to the fact that the algorithm relies on registers and shared-memory per block to speed up the solution and hence at 4096 it runs out of resources per block.  
Details of the algorithm is given in Algorithm \ref{alg:PCR}.

\begin{algorithm}[h!]
	\caption{Parallel Cyclic Reduction (PCR)}
	\label{alg:PCR}
	\begin{algorithmic}
		\State {define $a$ : Array storing the lower diagonal elements}
		\State {define $c$ : Array storing the lower diagonal elements}
		\State {define $a_{old}$ : Container to save old values of $a$}
		\State {define $c_{old}$ : Container to save old values of $c$}
		\State {define $x$ : Array storing the solution to the system}
		\State {define $r$ : local variable}
		\For{ $i=0, ..., P$} 		
		\State{$s=2^{i}$}
		\For{ $i=M_l-1: 0$} 		
		\State{$r=1./({1.-a_{old}[j] c_{old}[j-s]-c_{old}[j] a_{old}[j+1]}$})
		\State{$a[j]=-r \times (a_{old}[j] a_{old}[j-s])$}
		\State{$c[j]=-r \times (c_{old}[j] c_{old}[j+s])$}
		\State{$x[j]= r \times (x_{old}[j]-a_{old}[j] x_{old}[j-s]-c_{old}[j] d_{old}[j+s])$}
		\EndFor		
		\EndFor		
		\State{Return $x$}
	\end{algorithmic}
\end{algorithm}

\subsection{cuSPARSE}
NVIDIA offers \texttt{cuSPARSE} that contains a set of basic algorithms for sparse linear systems for GPUs. 
A comparison of several methods included in \texttt{cuSPARSE} is given in \cite{zhang2010fast}. NVIDIA offers CR, PCR and hybrid CR-PCR algorithms in \texttt{cuSPARSE} library \cite{naumov2010cusparse}. They also have pivoting versions.
\texttt{cuSPARSE} has a tridiagonal solver that can perform solves in batch mode with multiple right hand sides which is very promising for the system under consideration in \texttt{PittPack}. The only issue is that the systems obtained after the two FFT transformations have different diagonal elements and hence, the batch solve in \texttt{cuSPARSE} may not be utilized in \texttt{PittPack} efficiently. The only option is to solve every system separately, which has two disadvantages. First, the tridiagonal solver library provides the functions that are global and have to be called from the host. For a batch system of $nxChunk \times nyChunk$ the overhead of calling the host function repeatedly grows linearly. Second, we lose batch parallelism that significantly affects the prformance. For these reasons we do not consider \texttt{cuSPARSE} in \texttt{PittPack}.

\subsection{Comparison of Tridiagonal Solvers}
 We consider two test cases to quantify the impact of the tridiagonal solvers on the overall performance to solve the three dimensional Poisson's equation. In both test cases, we continually increase the mesh size in $z$-direction and record the execution times for each method using a single NVIDIA Tesla V100 GPU. Regardless of the number of GPUs, tridiagonal system solutions are performed locally at each GPU. We chose a single GPU to find an upper bound for the effect of tridiagonal solver on the overall Poisson solve. In both cases,  we continually increase the mesh size in $z-$direction and record the execution times using each solver.  

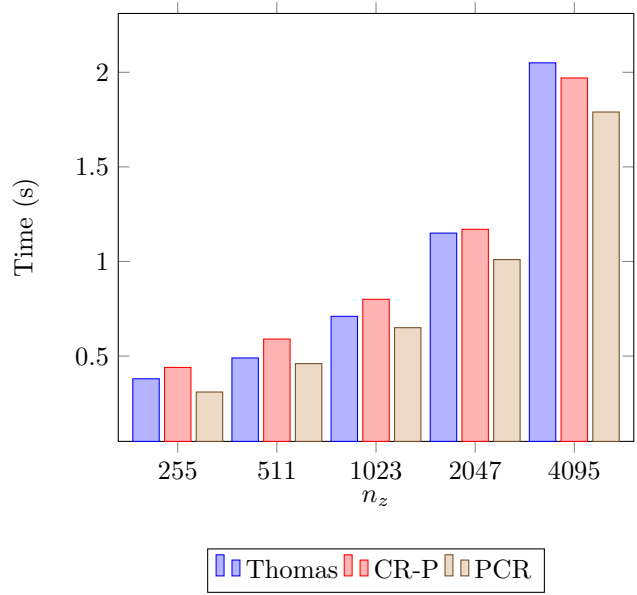
\begin{figure}[h!]
    \centering
    \begin{tikzpicture}
    \begin{axis}[
    ybar,
    enlargelimits=0.15,
    legend style={at={(0.5,-0.25)},
      anchor=north,legend columns=-1},
    ylabel={Time (s)},
    symbolic x coords={255,511,1023,2047,4095,},
    xtick=data,
    xlabel={$n_z$},
    nodes near coords align,
    nodes near coords align={vertical},
    ]
\addplot coordinates {(255,0.38)(511,0.49) (1023,0.71) (2047,1.15) (4095,2.05)};
\addplot coordinates {(255,0.44) (511,0.59) (1023,0.8) (2047,1.17) (4095,1.97)};
\addplot coordinates {(255,0.31)(511,0.46) (1023,0.65) (2047,1.01) (4095,1.79)};
\legend{Thomas, CR-P, PCR}
\end{axis}
\end{tikzpicture}
\caption{Comparison of the total runtime of the solution of the Poisson equation  using different solution methods for the tridiagonal matrix solution in \texttt{PittPack}. NVIDIA V100 GPU is used. \texttt{nxChunk=nyChunk=64}} 
    \label{fig:compare_three_methods}
\end{figure}

\begin{figure}[h!]
    \centering
    \begin{tikzpicture}
    \begin{axis}[
    ybar,
    enlargelimits=0.15,
    legend style={at={(0.5,-0.25)},
      anchor=north,legend columns=-1},
    ylabel={Time (s)},
    symbolic x coords={8191,16383,32767,65353},
    xtick=data,
    xlabel={$n_z$},
    nodes near coords align,
    nodes near coords align={vertical},
    ]
\addplot coordinates {(8191,3.12)(16383,6.00) (32767,11.08) (65353,23.37)};
\addplot coordinates {(8191,3.82)(16383,7.47) (32767,14.85) (65353,29.64)};
\legend{Thomas, CR-P}
\end{axis}
\end{tikzpicture}
\caption{Total run\-time of the Poissons's solve using Thomas and CR-P algorithms on a single NVIDIA V100 GPU,
\texttt{nxChunk=nyChunk=64}}
    \label{fig:compare_two_methods}
\end{figure}
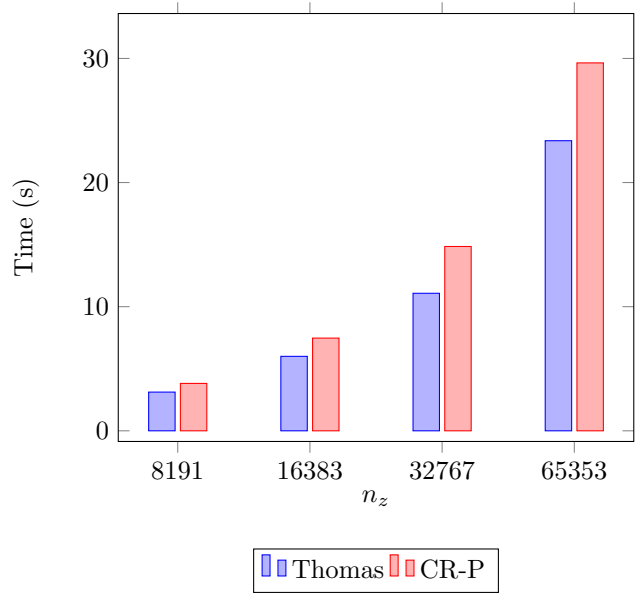

The first test case accommodates the restriction of the mesh size in the $z-$direction on the PCR algorithm, and hence limits the mesh size to 4096 in that direction. The results for this analysis is given in Fig. \ref{fig:compare_three_methods}. As expected, PCR algorithm is the fastest of all and CR-P and Thomas follow that method, respectively. Quantitatively, for the mesh size of 4095, PCR is 9.1\% faster than CR-P and 12.7\% faster than the memory-efficient implementation of the Thomas's method.
In the second test case, 

The second test case, compares CR-P with Thomas algorithm for a moderate mesh size in the $z-$direction. As we have mentioned earlier, PCR can not go beyond 4096 points and, therefore, excluded from this analysis. CR-P algorithm for all test cases is faster than Thomas. We note here that the 2D batch-solution mode is used for all the solutions and the effect of the tridiagonal matrix solution on the overall run-time is presented in Fig. \ref{fig:compare_two_methods}. The biggest mesh sizes considered in this study has the dimensions of $16384^3$. 
In this range, CR-P only offers a minor improvement in execution time over the low memory Thomas algorithm.
Therefore, we adopt the Thomas algorithm for the scaling analysis of \texttt{PittPack}. 

\section{Distributed-Memory Considerations for FFT}
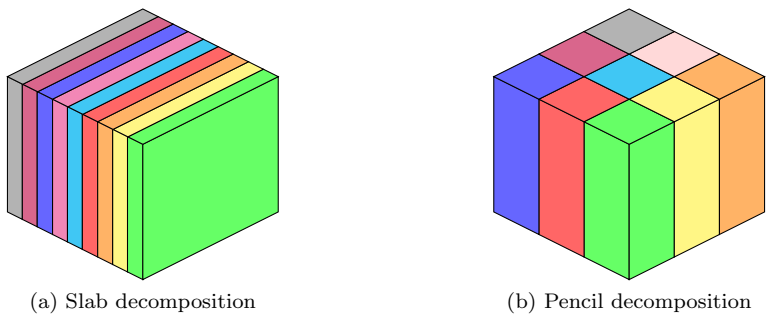
\begin{figure}[h!]
\centering
\subfloat[Slab decomposition]{
\begin{tikzpicture}[every node/.style={minimum size=1cm},on grid, scale=0.6]
\begin{scope}[every node/.append style={yslant=-0.5},yslant=-0.5]
   
   \filldraw[fill=gray!60!white] (0,0) rectangle (0.33,3);
   \filldraw[fill=purple!60!white] (0.33,0) rectangle (0.66,3);
   \filldraw[fill=blue!60!white] (0.66,0) rectangle (1,3);
   
   \filldraw[fill=magenta!60!white]  (1,0) rectangle (1.33,3);
   \filldraw[fill=cyan!60!white]  (1.33,0) rectangle (1.66,3);
   \filldraw[fill=red!60!white]  (1.66,0) rectangle (2.,3);
   
   \filldraw[fill=green!60!white]  (3,0) rectangle (2.66,3);
   \filldraw[fill=yellow!60!white]  (2.66,0) rectangle (2.33,3);
   \filldraw[fill=orange!60!white]  (2.33,0) rectangle (2.0,3);
   \end{scope}
\begin{scope}[every node/.append style={yslant=0.5},yslant=0.5]
  \filldraw[fill=green!60!white] (3,-3) rectangle +(3,3);
\end{scope}
\begin{scope}[every node/.append style={
    yslant=0.5,xslant=-1},yslant=0.5,xslant=-1
  ]
  \shade[bottom color=white!70, top color=white!70] (6,3) rectangle +(-3,-3);
   
   \filldraw[fill=green!60!white] (3,0) rectangle (6,0.33);
   \filldraw[fill=yellow!60!white] (3,.33) rectangle (6,0.66);
   \filldraw[fill=orange!60!white] (3,0.66) rectangle (6,1);
   
   \filldraw[fill=red!60!white] (3,1.) rectangle (6,1.33);
   \filldraw[fill=cyan!60!white] (3,1.33) rectangle (6,1.66);
   \filldraw[fill=magenta!60!white] (3,1.66) rectangle (6,2);
  
   \filldraw[fill=blue!60!white] (3,2) rectangle (6,2.33);
   \filldraw[fill=purple!60!white] (3, 2.33) rectangle (6,2.66);
   \filldraw[fill=gray!60!white] (3,2.66) rectangle (6,3);

\end{scope}
\end{tikzpicture}
\label{subfig:slab}
} 
\hspace{2.5 cm}
\subfloat[Pencil decomposition]{
\begin{tikzpicture}[every node/.style={minimum size=1cm},on grid, scale=0.6]
\begin{scope}[every node/.append style={yslant=-0.5},yslant=-0.5]
   \filldraw[fill=green!60!white] (3,0) rectangle (2,3);
   \filldraw[fill=red!60!white]  (1,0) rectangle (2,3);
   \filldraw[fill=blue!60!white] (0,0) rectangle (1,3);
\end{scope}
\begin{scope}[every node/.append style={yslant=0.5},yslant=0.5]
  
  \filldraw[fill=green!60!white] (3, -3) rectangle(4,0) ;
  \filldraw[fill=yellow!60!white] (4, -3) rectangle(5,0) ;
  \filldraw[fill=orange!60!white] (5, -3) rectangle(6,0) ;
\end{scope}
\begin{scope}[every node/.append style={
    yslant=0.5,xslant=-1},yslant=0.5,xslant=-1
  ]
  \shade[bottom color=white!70, top color=white!70] (6,3) rectangle +(-3,-3);
   \filldraw[fill=green!60!white] (3,0) rectangle (4,1);
   \filldraw[fill=red!60!white] (3,1) rectangle (4,2);
   \filldraw[fill=blue!60!white] (3,2) rectangle (4,3);
   
   \filldraw[fill=yellow!60!white] (4,0) rectangle(5,1);
   \filldraw[fill=cyan!60!white] (4,1) rectangle (5,2);
   \filldraw[fill=purple!60!white] (4,2) rectangle (5,3);
   
   \filldraw[fill=orange!60!white] (5,0) rectangle(6,1);
   \filldraw[fill=pink!60!white] (5,1) rectangle (6,2);
   \filldraw[fill=gray!60!white] (5,2) rectangle (6,3);
   
\end{scope}
\end{tikzpicture}
\label{subfig:pencil}
}
\caption{Pencil and slab decomposition methods.}
\label{fig:slabPencil}
\end{figure}

Performing Fourier transform in parallel is not a trivial task. FFT needs to have access to the entire signal before calculating the transformations. Consequently, data migration and domain decomposition are important decisions affecting parallel performance of FFT. There are two approaches for domain decomposition: i) slab decomposition; ii) pencil decomposition. These decompositions are graphically illustrated in Figs. \ref{subfig:slab} and \ref{subfig:pencil}, respectively. 
Slab decomposition (see Fig. \ref{subfig:slab}) relies on the fact that each processor has local access to the entire signal in one direction to perform 2D decomposition.
However, this approach does not have good scalability characteristics due to its high surface-to-volume ratio \cite{foster1995designing}.  
Subscribing more resources to a fixed problem reduces computation per process while keeping the communication size constant, which eventually degrades scalability.

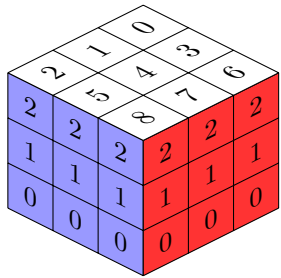
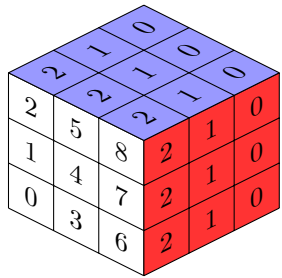
\begin{figure}[h!]
\subfloat[arrangement in Z-direction]
{
\centering
\begin{tikzpicture}[every node/.style={minimum size=1cm},on grid, scale=0.6]
\begin{scope}[every node/.append style={yslant=-0.5},yslant=-0.5]
  \shade[right color=blue!40, left color=blue!40] (0,0) rectangle +(3,3);
  \node at (0.5,2.5) {2};
  \node at (1.5,2.5) {2};
  \node at (2.5,2.5) {2};
  \node at (0.5,1.5) {1};
  \node at (1.5,1.5) {1};
  \node at (2.5,1.5) {1};
  \node at (0.5,0.5) {0};
  \node at (1.5,0.5) {0}; 
  \node at (2.5,0.5) {0};
  \draw (0,0) grid (3,3);
\end{scope}

\begin{scope}[every node/.append style={yslant=0.5},yslant=0.5]
  \shade[right color=red!80,left color=red!80] (3,-3) rectangle +(3,3);
  \node at (3.5,-0.5) {2};
  \node at (4.5,-0.5) {2};
 \node at (5.5,-0.5) {2};
 \node at (3.5,-1.5) {1};
  \node at (4.5,-1.5) {1};
  \node at (5.5,-1.5) {1};
  \node at (3.5,-2.5) {0};
  \node at (4.5,-2.5) {0};
  \node at (5.5,-2.5) {0};
  \draw (3,-3) grid (6,0);
\end{scope}

\begin{scope}[every node/.append style={
    yslant=0.5,xslant=-1},yslant=0.5,xslant=-1
  ]
  \shade[bottom color=white!70, top color=white!70] (6,3) rectangle +(-3,-3);
  \node at (3.5,2.5) {2};
  \node at (3.5,1.5) {5};
  \node at (3.5,0.5) {8};
  \node at (4.5,2.5) {1};
  \node at (4.5,1.5) {4};
  \node at (4.5,0.5) {7};
  \node at (5.5,2.5) {0};
  \node at (5.5,1.5) {3};
  \node at (5.5,0.5) {6};
  \draw (3,0) grid (6,3);
\end{scope}
\end{tikzpicture}
\label{fig:decomp0}
}
\hspace{1.cm}
\subfloat[arrangement in X-direction after data-transfer]{
\centering
\begin{tikzpicture}[every node/.style={minimum size=1cm},on grid, scale=0.6]
\begin{scope}[every node/.append style={yslant=-0.5},yslant=-0.5]
  \shade[bottom color=white!70, top color=white!70] (0,0) rectangle +(3,3);
  \node at (0.5,2.5) {2};
  \node at (1.5,2.5) {5};
  \node at (2.5,2.5) {8};
  \node at (0.5,1.5) {1};
  \node at (1.5,1.5) {4};
  \node at (2.5,1.5) {7};
  \node at (0.5,0.5) {0};
  \node at (1.5,0.5) {3}; 
  \node at (2.5,0.5) {6};
  \draw (0,0) grid (3,3);
\end{scope}

\begin{scope}[every node/.append style={yslant=0.5},yslant=0.5]
  \shade[right color=red!80,left color=red!80] (3,-3) rectangle +(3,3);
  \node at (3.5,-0.5) {2};
  \node at (4.5,-0.5) {1};
 \node at (5.5,-0.5) {0};
 \node at (3.5,-1.5) {2};
  \node at (4.5,-1.5) {1};
  \node at (5.5,-1.5) {0};
  \node at (3.5,-2.5) {2};
  \node at (4.5,-2.5) {1};
  \node at (5.5,-2.5) {0};
  \draw (3,-3) grid (6,0);
\end{scope}

\begin{scope}[every node/.append style={
    yslant=0.5,xslant=-1},yslant=0.5,xslant=-1
  ]
  \shade[right color=blue!40, left color=blue!40] (6,3) rectangle +(-3,-3);
  \node at (3.5,2.5) {2};
  \node at (3.5,1.5) {2};
  \node at (3.5,0.5) {2};
  \node at (4.5,2.5) {1};
  \node at (4.5,1.5) {1};
  \node at (4.5,0.5) {1};
  \node at (5.5,2.5) {0};
  \node at (5.5,1.5) {0};
  \node at (5.5,0.5) {0};
  \draw (3,0) grid (6,3);
\end{scope}
\end{tikzpicture}

\label{fig:decomp1}
}
\caption{Chunked-pencil decomposition. White surface tags show the processor rank. Numbers on the sides faces colored in blue and red depict chunk ID in each processor.}
\label{fig:chunkedPencil}
\end{figure}

The second approach, adopted in \texttt{PittPack}, is the pencil decomposition as shown in Fig. \ref{subfig:pencil}. In this approach, contrary to the slab decomposition, subscribing more resources, leads to reduction in message size, and hence it will perform better in terms of scalability. 
Most of the open-source libraries (e.g., \texttt{P3DFFT} \cite{pekurovsky2012p3dfft} and \texttt{PFFT} \cite{pippig2013pfft}) adopt the pencil decomposition because of its superior scalability characteristics.
There are several publications that discuss the pencil decomposition strategy \cite{li20102decomp, pippig2013pfft}.
\texttt{PittPack} adopts chunked-pencil decomposition as shown in Fig. \ref{fig:chunkedPencil}. This decomposition is mainly designed to enhance data movement and nodal communications. It also helps design a custom communication pattern with smaller memory footprint. 

\texttt{MPI} provides a collective communication function to perform the data exchange in the pencil decomposition. Most CPU-based FFT solvers rely on the efficiency of \texttt{MPI\_Alltoall} (or \texttt{MPI\_Alltoallv}) to make one call instead of several \texttt{MPI\_Isend} and \texttt{MPI\_Irecv} at each redistribution phase. For example, \texttt{2DDECOMP} as well as \texttt{P3DFFT} both take advantage of the this function to perform nodal communications. 
As discussed in \texttt{MPI} on Million Processor \cite{balaji2009mpi}, an all-to-all communication in an algorithm can hinder scalability. Actually, all-to-all communications can be avoided in parallel FFT. Because \texttt{MPI\_Alltoall} does not send information to every process, but rather fills some of the destination entries with zeros. Although most \texttt{MPI} implementations investigate this pattern and and only communicate with processes that have non-zero data, they still have to scan
through the entire array of data to obtain this information. This is another burden on the user, as the user needs to allocate and initialize this array. For extreme scale computing, any operation that grows linearly with the number of processes is not desirable and should be avoided if possible.  \texttt{MPI} 3.0 provides new functions to improve upon this bottleneck, which is referred to as neighborhood collectives or sepcifically \texttt{MPI\_Ineighbor\_alltoall} function.

In \texttt{MPI\_Ineighbor\_alltoall}, collective operation only occurs in a graph created before calling the collective communication function. Although, some of the collective operations can use \texttt{MPI\_IN\_PLACE}, and, potentially, save on memory consumption depending on the details of implementation, such an implementation would eliminate the opportunity for overlapping communication and data transfer via PCI-bus in a heterogeneous cluster with accelerators. This issue was also pointed out by the developers of \emph{AccFFT} \cite{gholami2015accfft}.
\texttt{PittPack} implements the neighborhood collective function in a customized fashion using the \textit{chunked-array layout} for storage of the data as illustrated in Fig. \ref{fig:chunkedPencil}. 
We utilize \texttt{MPI\_Dist\_graph\_create\_adjacent()} to generate the nodal connectivity in $x$ and $y$ directions. 
Assuming there are same number of processes in $x$ and $y$ directions, we then call the \texttt{MPI\_Ineighbor\_alltoall} 
function to perform communication only in the neighborhood.
Note that utilizing neighborhood collectives in the pencil decomposition, the number of elements in the connectivity array grows with the square root of the processors assuming that $p_{x}=p_{y}=p_{0}$ ($p_{x} $ and $p_{y}$ are the number of tasks in $x$ and $y$ directions), as opposed to linear when regular \texttt{MPI\_alltoall} is used. Therefore, the new neighborhood collective function has better scalability that the regular \texttt{MPI\_alltoall}.
We implemented \texttt{MPI\_Ineighbor\_alltoall} method in \texttt{PittPack}. One can choose this method by setting \texttt{COMM\_PATTERN=1} in the configuration file. 
While, the new functionality of \texttt{MPI} is ideal for CPU only communications, it is not very efficient when it comes to heterogenous computing. As mentioned previously a key point in circumventing the slower PCI bus data tranfer, is to overlap nodal communications with the data transfer. This requires custom communication pattern design. 
In this connection, \texttt{PittPack} offers two more efficient communication patterns to enable message passing and data-copy overlap in the next section.

\subsection{Pairwise Exchange Communication Pattern}
\begin{figure}[h!]
\centering
\usetikzlibrary{arrows}
\pagestyle{empty}
\definecolor{xfqqff}{rgb}{0.4980392156862745,0,1}
\definecolor{qqccqq}{rgb}{0,0.8,0}
\definecolor{ffqqff}{rgb}{1,0,1}
\definecolor{ttqqqq}{rgb}{0.2,0,0}
\definecolor{ffqqqq}{rgb}{1,0,0}
\subfloat[Stage I: each processor sends its (i+1)th chunk to the temporary container shown as a circle.]{
\begin{tikzpicture}[line cap=round,line join=round,>=triangle 45,x=1cm,y=1cm,scale=0.6]
\clip(-.3495719145699936,-1.691431805043052) rectangle (14.05013465864896,8.38742144305839);
\fill[line width=2pt,color=ffqqqq,fill=ffqqqq,fill opacity=0.81] (2,0) -- (3.5,0) -- (3.5,1.5) -- (2,1.5) -- cycle;
\fill[line width=2pt,color=ffqqqq,fill=ffqqqq,fill opacity=0.81] (2,1.5) -- (3.5,1.5) -- (3.5,3) -- (2,3) -- cycle;
\fill[line width=2pt,color=ffqqqq,fill=ffqqqq,fill opacity=0.81] (2,3) -- (3.5,3) -- (3.5,4.5) -- (2,4.5) -- cycle;
\fill[line width=2pt,color=ffqqqq,fill=ffqqqq,fill opacity=0.81] (2,4.5) -- (3.5,4.5) -- (3.5,6) -- (2,6) -- cycle;
\fill[line width=2pt,color=ffqqff,fill=ffqqff,fill opacity=0.55] (5.5,0) -- (7,0) -- (7,1.5) -- (5.5,1.5) -- cycle;
\fill[line width=2pt,color=ffqqff,fill=ffqqff,fill opacity=0.55] (5.5,1.5) -- (7,1.5) -- (7,3) -- (5.5,3) -- cycle;
\fill[line width=2pt,color=ffqqff,fill=ffqqff,fill opacity=0.55] (5.5,3) -- (7,3) -- (7,4.5) -- (5.5,4.5) -- cycle;
\fill[line width=2pt,color=ffqqff,fill=ffqqff,fill opacity=0.55] (5.5,4.5) -- (7,4.5) -- (7,6) -- (5.5,6) -- cycle;
\fill[line width=2pt,color=qqccqq,fill=qqccqq,fill opacity=1] (9,0) -- (10.5,0) -- (10.5,1.5) -- (9,1.5) -- cycle;
\fill[line width=2pt,color=qqccqq,fill=qqccqq,fill opacity=1] (9,1.5) -- (10.5,1.5) -- (10.5,3) -- (9,3) -- cycle;
\fill[line width=2pt,color=qqccqq,fill=qqccqq,fill opacity=1] (9,3) -- (10.5,3) -- (10.5,4.5) -- (9,4.5) -- cycle;
\fill[line width=2pt,color=qqccqq,fill=qqccqq,fill opacity=1] (9,4.5) -- (10.5,4.5) -- (10.5,6) -- (9,6) -- cycle;
\fill[line width=2pt,color=xfqqff,fill=xfqqff,fill opacity=0.63] (12.5,0) -- (14,0) -- (14,1.5) -- (12.5,1.5) -- cycle;
\fill[line width=2pt,color=xfqqff,fill=xfqqff,fill opacity=0.63] (12.5,1.5) -- (14,1.5) -- (14,3) -- (12.5,3) -- cycle;
\fill[line width=2pt,color=xfqqff,fill=xfqqff,fill opacity=0.63] (12.5,3) -- (14,3) -- (14,4.5) -- (12.5,4.5) -- cycle;
\fill[line width=2pt,color=xfqqff,fill=xfqqff,fill opacity=0.63] (12.5,4.5) -- (14,4.5) -- (14,6) -- (12.5,6) -- cycle;
\draw [line width=2pt,color=ttqqqq] (2,0)-- (3.5,0);
\draw [line width=2pt,color=ttqqqq] (3.5,0)-- (3.5,1.5);
\draw [line width=2pt,color=ttqqqq] (3.5,1.5)-- (2,1.5);
\draw [line width=2pt,color=ttqqqq] (2,1.5)-- (2,0);
\draw [line width=2pt,color=ffqqqq] (2,1.5)-- (3.5,1.5);
\draw [line width=2pt,color=ttqqqq] (3.5,1.5)-- (3.5,3);
\draw [line width=2pt,color=ttqqqq] (2,3)-- (2,1.5);
\draw [line width=2pt,color=ffqqqq] (2,3)-- (3.5,3);
\draw [line width=2pt,color=ttqqqq] (3.5,3)-- (3.5,4.5);
\draw [line width=2pt,color=ttqqqq] (2,4.5)-- (2,3);
\draw [line width=2pt,color=ttqqqq] (3.5,4.5)-- (3.5,6);
\draw [line width=2pt,color=ttqqqq] (3.5,6)-- (2,6);
\draw [line width=2pt,color=ttqqqq] (2,6)-- (2,4.5);
\draw [line width=2pt] (1,3)  circle (0.5cm);
\draw [line width=2pt] (4.5,3) circle (0.5cm);
\draw [line width=2pt,color=ttqqqq] (5.5,0)-- (7,0);
\draw [line width=2pt,color=ttqqqq] (7,0)-- (7,1.5);
\draw [line width=2pt,color=ttqqqq] (7,1.5)-- (5.5,1.5);
\draw [line width=2pt,color=ttqqqq] (5.5,1.5)-- (5.5,0);
\draw [line width=2pt,color=ffqqff] (5.5,1.5)-- (7,1.5);
\draw [line width=2pt,color=ttqqqq] (7,1.5)-- (7,3);
\draw [line width=2pt,color=ttqqqq] (7,3)-- (5.5,3);
\draw [line width=2pt,color=ttqqqq] (5.5,3)-- (5.5,1.5);
\draw [line width=2pt,color=ffqqff] (5.5,3)-- (7,3);
\draw [line width=2pt,color=ttqqqq] (7,3)-- (7,4.5);
\draw [line width=2pt,color=ttqqqq] (7,4.5)-- (5.5,4.5);
\draw [line width=2pt,color=ttqqqq] (5.5,4.5)-- (5.5,3);
\draw [line width=2pt,color=ffqqff] (5.5,4.5)-- (7,4.5);
\draw [line width=2pt,color=ttqqqq] (7,4.5)-- (7,6);
\draw [line width=2pt,color=ttqqqq] (7,6)-- (5.5,6);
\draw [line width=2pt,color=ttqqqq] (5.5,6)-- (5.5,4.5);
\draw [line width=2pt] (8,3) circle (0.4871938915987716cm);
\draw [line width=2pt,color=ttqqqq] (9,0)-- (10.5,0);
\draw [line width=2pt,color=ttqqqq] (10.5,0)-- (10.5,1.5);
\draw [line width=2pt,color=ttqqqq] (10.5,1.5)-- (9,1.5);
\draw [line width=2pt,color=ttqqqq] (9,1.5)-- (9,0);
\draw [line width=2pt,color=qqccqq] (9,1.5)-- (10.5,1.5);
\draw [line width=2pt,color=ttqqqq] (10.5,1.5)-- (10.5,3);
\draw [line width=2pt,color=ttqqqq] (10.5,3)-- (9,3);
\draw [line width=2pt,color=ttqqqq] (9,3)-- (9,1.5);
\draw [line width=2pt,color=qqccqq] (9,3)-- (10.5,3);
\draw [line width=2pt,color=ttqqqq] (10.5,3)-- (10.5,4.5);
\draw [line width=2pt,color=ttqqqq] (10.5,4.5)-- (9,4.5);
\draw [line width=2pt,color=ttqqqq] (9,4.5)-- (9,3);
\draw [line width=2pt,color=qqccqq] (9,4.5)-- (10.5,4.5);
\draw [line width=2pt,color=ttqqqq] (10.5,4.5)-- (10.5,6);
\draw [line width=2pt,color=ttqqqq] (10.5,6)-- (9,6);
\draw [line width=2pt,color=ttqqqq] (9,6)-- (9,4.5);
\draw [line width=2pt] (11.5,3) circle (0.5cm);
\draw [line width=2pt,color=ttqqqq] (14,0)-- (14,1.5);
\draw [line width=2pt,color=ttqqqq] (14,1.5)-- (12.5,1.5);
\draw [line width=2pt,color=ttqqqq] (12.5,1.5)-- (12.5,0);
\draw [line width=2pt,color=xfqqff] (12.5,1.5)-- (14,1.5);
\draw [line width=2pt,color=ttqqqq] (14,1.5)-- (14,3);
\draw [line width=2pt,color=ttqqqq] (14,3)-- (12.5,3);
\draw [line width=2pt,color=ttqqqq] (12.5,3)-- (12.5,1.5);
\draw [line width=2pt,color=xfqqff] (12.5,3)-- (14,3);
\draw [line width=2pt,color=ttqqqq] (14,3)-- (14,4.5);
\draw [line width=2pt,color=xfqqff] (14,4.5)-- (12.5,4.5);
\draw [line width=2pt,color=ttqqqq] (12.5,4.5)-- (12.5,3);
\draw [line width=2pt,color=xfqqff] (12.5,4.5)-- (14,4.5);
\draw [line width=2pt,color=ttqqqq] (14,4.5)-- (14,6);
\draw [line width=2pt,color=ttqqqq] (14,6)-- (12.5,6);
\draw [line width=2pt,color=ttqqqq] (12.5,6)-- (12.5,4.5);
\draw [line width=2pt] (2,4.5)-- (3.5,4.5);
\draw [line width=2pt] (2,3)-- (3.5,3);
\draw [line width=2pt] (2,1.5)-- (3.5,1.5);
\draw [line width=2pt] (5.5,1.5)-- (7,1.5);
\draw [line width=2pt] (5.5,3)-- (7,3);
\draw [line width=2pt] (7,4.5)-- (5.5,4.5);
\draw [line width=2pt] (9,1.5)-- (10.5,1.5);
\draw [line width=2pt] (10.5,3)-- (9,3);
\draw [line width=2pt] (9,4.5)-- (10.5,4.5);
\draw [line width=2pt] (12.5,1.5)-- (14,1.5);
\draw [line width=2pt] (12.5,3)-- (14,3);
\draw [line width=2pt] (12.5,4.5)-- (14,4.5);
\draw [->,line width=1pt] (3.5,2.2819092293177206) -- (4.274066311530069,2.553957436544031);
\draw [->,line width=1pt] (7,3.9687217994838075) -- (7.684497961234725,3.371236247618832);
\draw [->,line width=1pt] (10.5,5.419380609826642) -- (11.295407062289792,3.45622552519462);
\draw [line width=2pt] (5.5,3)-- (4.999995062785096,3.002221978966444);
\draw [line width=2pt] (2,3)-- (1.4999989875375679,2.998993788587347);
\draw [line width=2pt] (12.5,3)-- (11.999999892434143,3.0003279723230394);
\draw [line width=2pt] (9,3)-- (8.487188681156347,2.997746788004893);
\draw [line width=1pt] (13.239531603632546,0.0019116744993419713)-- (13.264284219722576,-1.273595684107295);
\draw [line width=1pt] (13.264284219722576,-1.273595684107295)-- (1.077012122472856,-1.3808150867223994);
\draw [->,line width=1pt] (1.077012122472856,-1.3808150867223994) -- (1.0256714248738255,2.5006594569385063);
\draw [line width=1pt] (12.5,0)-- (14,0);
\filldraw[draw=none,fill=yellow!180!white] (2.05,1.6) rectangle +(1.4,1.4);
\filldraw[draw=none,fill=yellow!180!white] (5.56,3.05) rectangle +(1.4,1.4);
\filldraw[draw=none,fill=yellow!180!white] (9.05,4.55) rectangle +(1.4,1.4);
\filldraw[draw=none,fill=yellow!180!white] (12.56,0.05) rectangle +(1.4,1.4);
\end{tikzpicture}}
\qquad
\subfloat[Stage II: each processor sends (i-1)th chunk to the destination Rank-1]{
\begin{tikzpicture}[line cap=round,line join=round,>=triangle 45,x=1cm,y=1cm,scale=0.6]
\clip(-.3495719145699936,-1.691431805043052) rectangle (14.05013465864896,8.38742144305839);
\fill[line width=2pt,color=ffqqqq,fill=ffqqqq,fill opacity=0.81] (2,0) -- (3.5,0) -- (3.5,1.5) -- (2,1.5) -- cycle;
\fill[line width=2pt,color=ffqqqq,fill=ffqqqq,fill opacity=0.81] (2,1.5) -- (3.5,1.5) -- (3.5,3) -- (2,3) -- cycle;
\fill[line width=2pt,color=ffqqqq,fill=ffqqqq,fill opacity=0.81] (2,3) -- (3.5,3) -- (3.5,4.5) -- (2,4.5) -- cycle;
\fill[line width=2pt,color=ffqqqq,fill=ffqqqq,fill opacity=0.81] (2,4.5) -- (3.5,4.5) -- (3.5,6) -- (2,6) -- cycle;
\fill[line width=2pt,color=ffqqff,fill=ffqqff,fill opacity=0.55] (5.5,0) -- (7,0) -- (7,1.5) -- (5.5,1.5) -- cycle;
\fill[line width=2pt,color=ffqqff,fill=ffqqff,fill opacity=0.55] (5.5,1.5) -- (7,1.5) -- (7,3) -- (5.5,3) -- cycle;
\fill[line width=2pt,color=ffqqff,fill=ffqqff,fill opacity=0.55] (5.5,3) -- (7,3) -- (7,4.5) -- (5.5,4.5) -- cycle;
\fill[line width=2pt,color=ffqqff,fill=ffqqff,fill opacity=0.55] (5.5,4.5) -- (7,4.5) -- (7,6) -- (5.5,6) -- cycle;
\fill[line width=2pt,color=qqccqq,fill=qqccqq,fill opacity=1] (9,0) -- (10.5,0) -- (10.5,1.5) -- (9,1.5) -- cycle;
\fill[line width=2pt,color=qqccqq,fill=qqccqq,fill opacity=1] (9,1.5) -- (10.5,1.5) -- (10.5,3) -- (9,3) -- cycle;
\fill[line width=2pt,color=qqccqq,fill=qqccqq,fill opacity=1] (9,3) -- (10.5,3) -- (10.5,4.5) -- (9,4.5) -- cycle;
\fill[line width=2pt,color=qqccqq,fill=qqccqq,fill opacity=1] (9,4.5) -- (10.5,4.5) -- (10.5,6) -- (9,6) -- cycle;
\fill[line width=2pt,color=xfqqff,fill=xfqqff,fill opacity=0.63] (12.5,0) -- (14,0) -- (14,1.5) -- (12.5,1.5) -- cycle;
\fill[line width=2pt,color=xfqqff,fill=xfqqff,fill opacity=0.63] (12.5,1.5) -- (14,1.5) -- (14,3) -- (12.5,3) -- cycle;
\fill[line width=2pt,color=xfqqff,fill=xfqqff,fill opacity=0.63] (12.5,3) -- (14,3) -- (14,4.5) -- (12.5,4.5) -- cycle;
\fill[line width=2pt,color=xfqqff,fill=xfqqff,fill opacity=0.63] (12.5,4.5) -- (14,4.5) -- (14,6) -- (12.5,6) -- cycle;
\draw [line width=2pt,color=ttqqqq] (2,0)-- (3.5,0);
\draw [line width=2pt,color=ttqqqq] (3.5,0)-- (3.5,1.5);
\draw [line width=2pt,color=ttqqqq] (3.5,1.5)-- (2,1.5);
\draw [line width=2pt,color=ttqqqq] (2,1.5)-- (2,0);
\draw [line width=2pt,color=ffqqqq] (2,1.5)-- (3.5,1.5);
\draw [line width=2pt,color=ttqqqq] (3.5,1.5)-- (3.5,3);
\draw [line width=2pt,color=ttqqqq] (2,3)-- (2,1.5);
\draw [line width=2pt,color=ffqqqq] (2,3)-- (3.5,3);
\draw [line width=2pt,color=ttqqqq] (3.5,3)-- (3.5,4.5);
\draw [line width=2pt,color=ttqqqq] (2,4.5)-- (2,3);
\draw [line width=2pt,color=ttqqqq] (3.5,4.5)-- (3.5,6);
\draw [line width=2pt,color=ttqqqq] (3.5,6)-- (2,6);
\draw [line width=2pt,color=ttqqqq] (2,6)-- (2,4.5);
\draw [line width=2pt,color=ttqqqq] (5.5,0)-- (7,0);
\draw [line width=2pt,color=ttqqqq] (7,0)-- (7,1.5);
\draw [line width=2pt,color=ttqqqq] (7,1.5)-- (5.5,1.5);
\draw [line width=2pt,color=ttqqqq] (5.5,1.5)-- (5.5,0);
\draw [line width=2pt,color=ffqqff] (5.5,1.5)-- (7,1.5);
\draw [line width=2pt,color=ttqqqq] (7,1.5)-- (7,3);
\draw [line width=2pt,color=ttqqqq] (7,3)-- (5.5,3);
\draw [line width=2pt,color=ttqqqq] (5.5,3)-- (5.5,1.5);
\draw [line width=2pt,color=ffqqff] (5.5,3)-- (7,3);
\draw [line width=2pt,color=ttqqqq] (7,3)-- (7,4.5);
\draw [line width=2pt,color=ttqqqq] (7,4.5)-- (5.5,4.5);
\draw [line width=2pt,color=ttqqqq] (5.5,4.5)-- (5.5,3);
\draw [line width=2pt,color=ffqqff] (5.5,4.5)-- (7,4.5);
\draw [line width=2pt,color=ttqqqq] (7,4.5)-- (7,6);
\draw [line width=2pt,color=ttqqqq] (7,6)-- (5.5,6);
\draw [line width=2pt,color=ttqqqq] (5.5,6)-- (5.5,4.5);
\draw [line width=2pt,color=ttqqqq] (9,0)-- (10.5,0);
\draw [line width=2pt,color=ttqqqq] (10.5,0)-- (10.5,1.5);
\draw [line width=2pt,color=ttqqqq] (10.5,1.5)-- (9,1.5);
\draw [line width=2pt,color=ttqqqq] (9,1.5)-- (9,0);
\draw [line width=2pt,color=qqccqq] (9,1.5)-- (10.5,1.5);
\draw [line width=2pt,color=ttqqqq] (10.5,1.5)-- (10.5,3);
\draw [line width=2pt,color=ttqqqq] (10.5,3)-- (9,3);
\draw [line width=2pt,color=ttqqqq] (9,3)-- (9,1.5);
\draw [line width=2pt,color=qqccqq] (9,3)-- (10.5,3);
\draw [line width=2pt,color=ttqqqq] (10.5,3)-- (10.5,4.5);
\draw [line width=2pt,color=ttqqqq] (10.5,4.5)-- (9,4.5);
\draw [line width=2pt,color=ttqqqq] (9,4.5)-- (9,3);
\draw [line width=2pt,color=qqccqq] (9,4.5)-- (10.5,4.5);
\draw [line width=2pt,color=ttqqqq] (10.5,4.5)-- (10.5,6);
\draw [line width=2pt,color=ttqqqq] (10.5,6)-- (9,6);
\draw [line width=2pt,color=ttqqqq] (9,6)-- (9,4.5);
\draw [line width=2pt,color=ttqqqq] (14,0)-- (14,1.5);
\draw [line width=2pt,color=ttqqqq] (14,1.5)-- (12.5,1.5);
\draw [line width=2pt,color=ttqqqq] (12.5,1.5)-- (12.5,0);
\draw [line width=2pt,color=xfqqff] (12.5,1.5)-- (14,1.5);
\draw [line width=2pt,color=ttqqqq] (14,1.5)-- (14,3);
\draw [line width=2pt,color=ttqqqq] (14,3)-- (12.5,3);
\draw [line width=2pt,color=ttqqqq] (12.5,3)-- (12.5,1.5);
\draw [line width=2pt,color=xfqqff] (12.5,3)-- (14,3);
\draw [line width=2pt,color=ttqqqq] (14,3)-- (14,4.5);
\draw [line width=2pt,color=xfqqff] (14,4.5)-- (12.5,4.5);
\draw [line width=2pt,color=ttqqqq] (12.5,4.5)-- (12.5,3);
\draw [line width=2pt,color=xfqqff] (12.5,4.5)-- (14,4.5);
\draw [line width=2pt,color=ttqqqq] (14,4.5)-- (14,6);
\draw [line width=2pt,color=ttqqqq] (14,6)-- (12.5,6);
\draw [line width=2pt,color=ttqqqq] (12.5,6)-- (12.5,4.5);
\draw [line width=2pt] (2,4.5)-- (3.5,4.5);
\draw [line width=2pt] (2,3)-- (3.5,3);
\draw [line width=2pt] (2,1.5)-- (3.5,1.5);
\draw [line width=2pt] (5.5,1.5)-- (7,1.5);
\draw [line width=2pt] (5.5,3)-- (7,3);
\draw [line width=2pt] (7,4.5)-- (5.5,4.5);
\draw [line width=2pt] (9,1.5)-- (10.5,1.5);
\draw [line width=2pt] (10.5,3)-- (9,3);
\draw [line width=2pt] (9,4.5)-- (10.5,4.5);
\draw [line width=2pt] (12.5,1.5)-- (14,1.5);
\draw [line width=2pt] (12.5,3)-- (14,3);
\draw [line width=2pt] (12.5,4.5)-- (14,4.5);
\draw [<-,line width=1pt] (3.5,2.2819092293177206) -- (5.474066311530069,0.553957436544031);
\draw [<-,line width=1pt] (7,3.9687217994838075) -- (8.9684497961234725,2.371236247618832);
\draw [<-,line width=1pt] (10.5,5.5819092293177206) -- (12.474066311530069,3.553957436544031);
\draw [<-,line width=1pt] (13.239531603632546,0.0019116744993419713)-- (13.264284219722576,-1.273595684107295);
\draw [line width=1pt] (13.264284219722576,-1.273595684107295)-- (1.077012122472856,-1.3808150867223994);
\draw [line width=1pt] (1.077012122472856,-1.3808150867223994) -- (1.0256714248738255,5.206594569385063);
\draw [line width=1pt] (1.0256714248738255,5.206594569385063) -- (2.0256714248738255,5.206594569385063);
\draw [line width=1pt] (12.5,0)-- (14,0);

\filldraw[draw=none,fill=yellow!180!white] (2.05,4.55) rectangle +(1.4,1.4);
\filldraw[draw=none,fill=yellow!180!white] (5.56,0.05) rectangle +(1.4,1.4);
\filldraw[draw=none,fill=yellow!180!white] (9.05,1.55) rectangle +(1.4,1.4);
\filldraw[draw=none,fill=yellow!180!white] (12.56,3.05) rectangle +(1.4,1.4);

\end{tikzpicture}}
\caption{Pairwise exchange communication pattern for chunked-pencil decomposition. }
\label{fig:commPattern}
\end{figure}
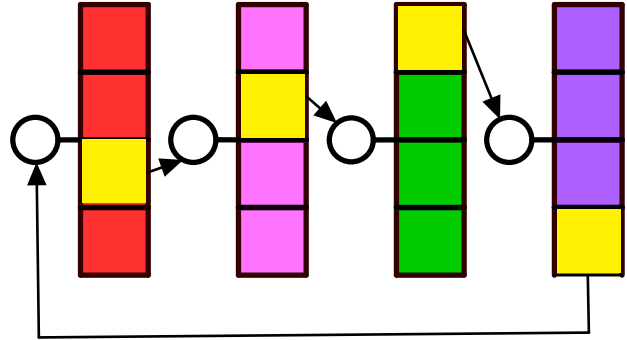
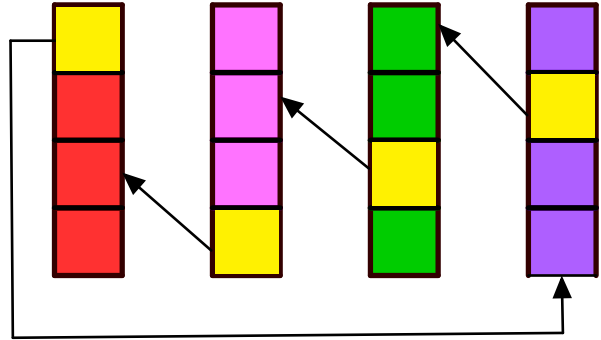

In \texttt{PittPack}, we utilize chunked-array decomposition and combine it with pairwise-exchange to achieve the overlap for intra-node communication and PCI-e data transfer. The chunked array improves data locality as all the messages are be contiguous. The algorithm assumes a periodic extension to the chunks. It is decomposed into two stages as follows:
\begin{enumerate}
\item \textit{Stage I.} Every processors sends \texttt{(rank+i)}th chunk to the processor with \texttt{rank+i}, where \texttt{i} is the loop count.
The data is received in an extra container of size \texttt{chunkSize}. In other words, the algorithm only uses one extra chunk of memory to perform the operations.
\item \textit{Stage II.} Every processor sends its \texttt{(rank-i)}th chunk to the processor with \texttt{(rank-i)}.
For all the indices that are smaller than zero or greater than number of chunks, the index is adjusted such that there
is a periodic extension from both ends. 
\end{enumerate}
This pattern is presented in Fig. \ref{fig:commPattern}.
Note that the buffer size required for this communication pattern is only \texttt{chunkSize}.
In other words, if we have $\texttt{nChunks} \times \texttt{chunkSize}$, this communication needs  $\texttt{(nChunks+1)} \times \texttt{chunkSize}$.

FFT on distributed memory platforms also requires a transpose operation. The method that we use accommodates
overlapping of the communication and computation to hide latency.
The user may choose this method by setting the \texttt{COMM\_PATTERN=0} in the configuration. We note that at the extreme scale, this communication pattern will not overwhelm the network, because one pair of send and receives are active at every stage only. 

\subsubsection{Fully Overlapped Communication Pattern}
The pairwise-exchange algorithm that was presented in the previous section is a memory-efficient approach for performing communication. If memory size is not an issue and maximum performance is desired, one may choose the fully overlap version in \texttt{PittPack} (\texttt{COMM\_PATTERN=2}) to gain a better performance. 
For this option, given an \texttt{nChunks*chunkSize} mesh, the required memory is \texttt{2*nChunks*chunkSize}.
Here, we explain this communication and data transfer strategy.

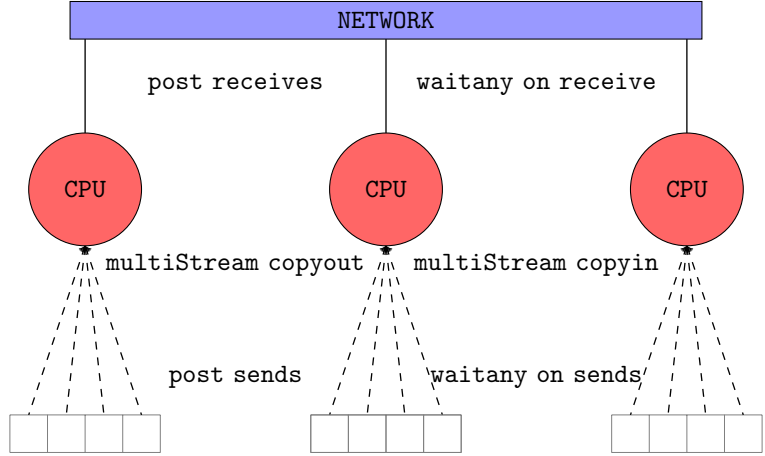
\begin{figure}[h!]
\centering
\begin{tikzpicture}
\filldraw[fill=red!60!white, draw=black]  ( -7, 0)  circle[radius=0.75] node{\texttt{CPU}};
\draw[step=0.5cm,gray,very thin,fill=green!60!white] (-8,-3) grid (-6,-3.5);
\draw[<-,line width=0.5pt,dashed] (-7.,-0.75) -- (-7.25,-3.0);
\draw[<-,line width=0.5pt,dashed] (-7.,-0.75) -- (-6.75,-3.0);
\draw[<-,line width=0.5pt,dashed] (-7.,-0.75) -- (-7.75,-3.0);
\draw[<-,line width=0.5pt,dashed] (-7.,-0.75) -- (-6.25,-3.0);

\filldraw[fill=red!60!white, draw=black]  ( -3, 0)  circle[radius=0.75] node{\texttt{CPU}};
\draw[step=0.5cm,gray,very thin] (-4,-3) grid (-2,-3.5);
\draw[<-,line width=0.5pt,dashed] (-3.,-0.75) -- (-3.75,-3.0);
\draw[<-,line width=0.5pt,dashed] (-3.,-0.75) -- (-3.25,-3.0);
\draw[<-,line width=0.5pt,dashed] (-3.,-0.75) -- (-2.75,-3.0);
\draw[<-,line width=0.5pt,dashed] (-3.,-0.75) -- (-2.25,-3.0);

\filldraw[fill=red!60!white, draw=black]  ( +1, 0)  circle[radius=0.75] node{\texttt{CPU}};
\draw[step=0.5cm,gray,very thin] (0,-3) grid (2,-3.5);
\draw[step=0.5cm,gray,very thin] (-4,-3) grid (-2,-3.5);
\draw[<-,line width=0.5pt,dashed] (1.,-0.75) -- (0.75,-3.0);
\draw[<-,line width=0.5pt,dashed] (1,-0.75) -- (1.25,-3.0);
\draw[<-,line width=0.5pt,dashed] (1,-0.75) -- (1.75,-3.0);
\draw[<-,line width=0.5pt,dashed] (1,-0.75) -- (0.25,-3.0);


\draw[line width=0.5pt] (-7.,2) -- (-7,0.75);
\draw[line width=0.5pt] (1.,2) -- (1 , 0.75);
\draw[line width=0.5pt] (-3,2) -- (-3 , 0.75);

\filldraw[fill=blue!40!white, draw=black] (-7.2,2) rectangle (1.2,2.5);
\draw[draw=none]  ( -3, 2.25) circle[radius=0.75] node{\texttt{NETWORK}};

\draw[draw=none]  ( -5, 1.4) circle[radius=0.75] node{\texttt{post} \texttt{receives}};
\draw[draw=none]  ( -5,-1.0) circle[radius=0.75] node{\texttt{multiStream} \texttt{copyout}};
\draw[draw=none]  ( -5, -2.5) circle[radius=0.75] node{\texttt{post} \texttt{sends}};

\draw[draw=none]  ( -1, 1.4) circle[radius=0.75] node{\texttt{waitany} \texttt{on} \texttt{receive}};
\draw[draw=none]  ( -1,-1.0) circle[radius=0.75] node{\texttt{multiStream} \texttt{copyin}};
\draw[draw=none]  ( -1, -2.5) circle[radius=0.75] node{\texttt{waitany} \texttt{on} \texttt{sends}};

\end{tikzpicture}
\caption{Fully overlapped communication pattern for chunked-pencil decomposition.}
\label{fig:gang}
\end{figure}

\definecolor{mygreen}{rgb}{0,0.6,0}
\definecolor{mygray}{rgb}{0.5,0.5,0.5}
\definecolor{mymauve}{rgb}{0.58,0,0.82}
\definecolor{backcolour}{rgb}{0.95,0.95,0.92}
\lstdefinestyle{customc}{
  belowcaptionskip=1\baselineskip, 
  breaklines=true,
  frame=L,
  xleftmargin=\parindent,
  language=C++,
  showstringspaces=false,
  backgroundcolor=\color{backcolour},
  basicstyle=\footnotesize\ttfamily,
  keywordstyle=\bfseries\color{red},
  commentstyle=\itshape\color{black},
  identifierstyle=\color{blue},
  stringstyle=\color{orange},
}

\begin{figure}[h!]
\lstset{style=customc}
\lstset{caption={Implementation of the fully overlapped communication pattern.},label=PittPackUtilize}
\lstset{basicstyle=\scriptsize}
\begin{center}
\begin{minipage}{\linewidth}
\begin{lstlisting}[frame=single]
.....
 // p0 is the number of neighbors including self
  for ( int i = 0; i <  p0 ; i++ )
    {
        // need to copy from device to host before sending
        ptr = &P( 0 ) + P.chunkSize * i;
        
        #pragma acc update self( ptr [0:end] ) async( i + 1 )
        MPI_Irecv( &R( 0 ) + R.chunkSize * i, R.chunkSize, MPI_DOUBLE, Nbrs[0][i], Nbrs[0][i], Comm, &( recv_request[i] ) );
    }
    for ( int i = 0; i < p0; i++ )
    {
// Wait on the given stream 
        acc_async_wait( i + 1 );
        
// Send the message after completion of the data transfer
        MPI_Isend( &P( 0 ) + P.chunkSize * i, P.chunkSize, MPI_DOUBLE, Nbrs[0][i], myRank, Comm, &( send_request[i] ) );
    }
    int indx;
    for ( int i = 0; i < p0; i++ )
    {
    // wait to make sure at least one message has arrived
        MPI_Waitany( p0, recv_request, &indx, recv_status );
        
    // record that index    
        indices[i] = indx;
        ptr        = &R( 0 ) + R.chunkSize * indx;
        #pragma acc update device( ptr [0:end] ) async( p0 + i + 1 )
    }
    int id;
    for ( int i = 0; i < p0; i++ )
    {
        id   = indices[i] + p0 + 1;
        indx = indices[i];
        acc_async_wait( id );
        MPI_Wait( &( send_request[indx] ), &( send_status[indx] ) );
        
        //overlap computation
        #pragma acc data present( P.P, R.P, this ) copyin( indx )
        #pragma acc parallel loop num_gangs( nGang )
        for ( int l = 0; l < P.chunkSize; l++ )
        {
            P( P.chunkSize * indx + l ) = R( R.chunkSize * indx + l );
        }
    }
\end{lstlisting}
\end{minipage}
\end{center}
\end{figure}
When solving the Poisson's equation for an extreme-scale problem on Oak Ridge National Laboratory's Titan supercomputer, we have observed that at 16834 GPUs the network throttles, i.e. there are too many requests for send and receive. This is expected as at any given time the number of send and receives per processors is increased proportional to the square root of number of \texttt{MPI} tasks. Note that network throttling did not occur when using the pairwise exchange communication pattern.

\newpage
\subsection{Data Shuffling}
It is preferred to perform FFT transform on a batches of data in a contiguous way.
In order to accomplish this, data needs to be rearranged before calling the \texttt{cuFFT} or \texttt{FFTW} 
functions.
The naive way of performing this operation is to use an auxiliary array of the same size. 
While this is the easiest and most parallelizable way in theory, it is not feasible for extreme scale computing.
Fortunately, since the data storage has a pattern and structure, it is possible to come up with an enhanced approach such that memory consumption is not excessive. Also, this strategy permits utilization of the shared memory for moderate block sizes.

\begin{figure}[h!]
\subfloat[Chunked array layout before performing FFT]{
\begin{tikzpicture}
\draw[step=1cm,gray] (-2,-2) grid (1,1);
\draw[draw=none] (-1.5,-1.5) circle[radius=0.4] node{$0$};
\filldraw[draw=none,fill=blue!40!white] (-0.5,-1.5) circle[radius=0.4] node{$1$};
\draw[draw=none] (+0.5,-1.5) circle[radius=0.2] node{$2$};
\filldraw[draw=none,fill=red!80!white] (-1.5,-0.5) circle[radius=0.4] node{$3$};
\draw[draw=none] (-0.5,-0.5) circle[radius=0.2] node{$4$};
\draw[draw=none] ( 0.5,-0.5) circle[radius=0.2] node{$5$};
\draw[draw=none] (-1.5,0.5) circle[radius=0.2] node{$6$};
\draw[draw=none] (-0.5,0.5) circle[radius=0.2] node{$7$};
\draw[draw=none] ( 0.5,0.5) circle[radius=0.2] node{$8$};

\draw[step=1cm,gray] (2,-2) grid (5,1);
\filldraw[draw=none,fill=green!40!white ] ( 2.5,-1.5) circle[radius=0.4] node{$9$};
\filldraw[draw=none,fill=yellow ] ( 2.5,-2.75) circle[radius=0.4] node{$tmp$};
\draw[draw=none] ( 3.5,-1.5) circle[radius=0.2] node{$10$};
\draw[draw=none] ( 4.5,-1.5) circle[radius=0.2] node{$11$};
\draw[draw=none] (2.5,-0.5) circle[radius=0.2] node{$12$};
\draw[draw=none] (3.5,-0.5) circle[radius=0.2] node{$13$};
\draw[draw=none] (4.5,-0.5) circle[radius=0.2] node{$14$};
\draw[draw=none] (2.5,0.5) circle[radius=0.2] node{$15$};
\draw[draw=none] (3.5,0.5) circle[radius=0.2] node{$16$};
\draw[draw=none] ( 4.5,0.5) circle[radius=0.2] node{$17$};

\draw[step=1cm,gray] (6,-2) grid (9,1);
\draw[draw=none] ( 6.5,-1.5) circle[radius=0.2] node{$18$};
\draw[draw=none] ( 7.5,-1.5) circle[radius=0.2] node{$19$};
\draw[draw=none] ( 8.5,-1.5) circle[radius=0.2] node{$20$};
\draw[draw=none] (6.5,-0.5) circle[radius=0.2] node{$21$};
\draw[draw=none] (7.5,-0.5) circle[radius=0.2] node{$22$};
\draw[draw=none] (8.5,-0.5) circle[radius=0.2] node{$23$};
\draw[draw=none] (6.5,0.5) circle[radius=0.2] node{$24$};
\draw[draw=none] (7.5,0.5) circle[radius=0.2] node{$25$};
\draw[draw=none] ( 8.5,0.5) circle[radius=0.2] node{$26$};

\draw[draw=none ] ( 2.5,-3.2) circle[radius=0.1];
\label{fig:current}
\end{tikzpicture}}
\qquad
\hspace{5mm}
\subfloat[Final arrangement before calling FFT]{
\begin{tikzpicture}
\draw[step=1cm,gray] (-2,-2) grid (1,1);
\draw[draw=none] (-1.5,-1.5) circle[radius=0.2] node{$0$};
\draw[draw=none] (-0.5,-1.5) circle[radius=0.2] node{$9$};
\draw[draw=none] (+0.5,-1.5) circle[radius=0.2] node{$18$};
\draw[draw=none] (-1.5,-0.5) circle[radius=0.2] node{$1$};
\draw[draw=none] (-0.5,-0.5) circle[radius=0.2] node{$10$};
\draw[draw=none] ( 0.5,-0.5) circle[radius=0.2] node{$19$};
\draw[draw=none] (-1.5,0.5) circle[radius=0.2] node{$2$};
\draw[draw=none] (-0.5,0.5) circle[radius=0.2] node{$11$};
\draw[draw=none] ( 0.5,0.5) circle[radius=0.2] node{$20$};

\draw[step=1cm,gray] (2,-2) grid (5,1);
\draw[draw=none] ( 2.5,-1.5) circle[radius=0.2] node{$3$};
\draw[draw=none] ( 3.5,-1.5) circle[radius=0.2] node{$12$};
\draw[draw=none] ( 4.5,-1.5) circle[radius=0.2] node{$21$};
\draw[draw=none] (2.5,-0.5) circle[radius=0.2] node{$4$};
\draw[draw=none] (3.5,-0.5) circle[radius=0.2] node{$13$};
\draw[draw=none] (4.5,-0.5) circle[radius=0.2] node{$22$};
\draw[draw=none] (2.5,0.5) circle[radius=0.2] node{$5$};
\draw[draw=none] (3.5,0.5) circle[radius=0.2] node{$14$};
\draw[draw=none] ( 4.5,0.5) circle[radius=0.2] node{$23$};

\draw[step=1cm,gray] (6,-2) grid (9,1);
\draw[draw=none] ( 6.5,-1.5) circle[radius=0.2] node{$6$};
\draw[draw=none] ( 7.5,-1.5) circle[radius=0.2] node{$15$};
\draw[draw=none] ( 8.5,-1.5) circle[radius=0.2] node{$24$};
\draw[draw=none] (6.5,-0.5) circle[radius=0.2] node{$7$};
\draw[draw=none] (7.5,-0.5) circle[radius=0.2] node{$16$};
\draw[draw=none] (8.5,-0.5) circle[radius=0.2] node{$25$};
\draw[draw=none] (6.5,0.5) circle[radius=0.2] node{$8$};
\draw[draw=none] (7.5,0.5) circle[radius=0.2] node{$17$};
\draw[draw=none] ( 8.5,0.5) circle[radius=0.2] node{$26$};
\draw[draw=none] ( 2.5,-2.75) circle[radius=0.1];
\label{fig:future}
\end{tikzpicture}
}

\caption{Illustration of memory efficient data shuffling among chunks before and after FFT.}
\label{fig:before}
\end{figure}
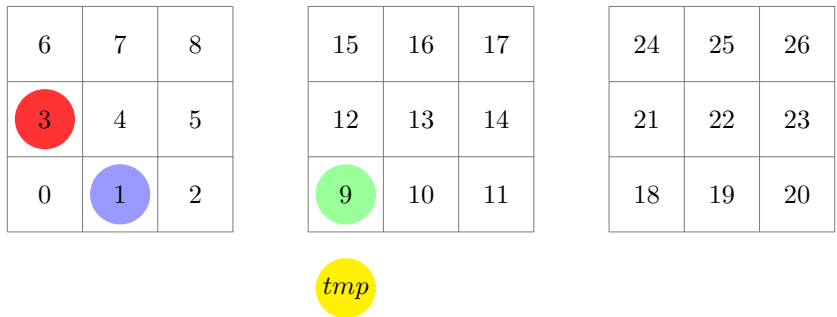
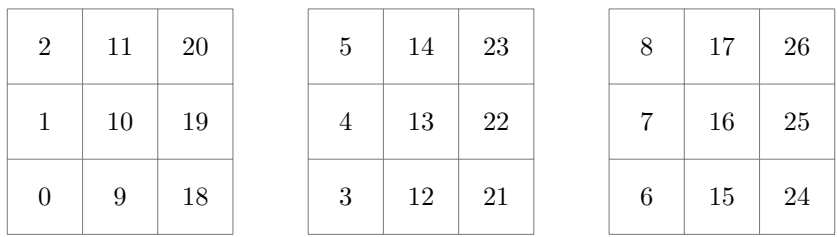

Here, we explain the method we have implemented in \texttt{PittPack}, which only uses an extra array of \texttt{nxChunk} (or \texttt{nyChunk}) depending on the direction of the transformation as an arbiter for swapping. 
Without loss of generality, we explain the algorithm using a \texttt{nChunk=nxChunk=nyChunk=3}. 
The front view of the grid with three chunks is presented in \ref{fig:before}.
Figure \ref{fig:current} demonstrates the current arrangement of the data. While Fig. \ref{fig:future} shows the desired location of the slots right before employing FFT transformation. We refer to the operation as \textit{data shuffling}. 
We emphasize here that block is an array of size \texttt{nxChunk} (or \texttt{nyChunk}) depending on the direction of interest. 

Starting with one element it is possible to follow the locations and trace the replacements back to the starting point. For instance, pick blocks No. 1, 3 and 9. The operations to place these blocks in their desired locations can be broken down as follows,   
\begin{enumerate}
\item move block No. 9 to the tmp array designated by yellow circle.
\item copy block No. 3 to the green spot
\item copy block No. 1 to the red spot
\item copy tmp array to blue spot
\end{enumerate} 

All these current and final destinations index calculations can be performed on the fly. 
The shuffling operations only requires an array of size \texttt{nxChunk} and hence, is very light weight.
This is very beneficial specifically when \texttt{nxChunk} is small enough to fit into the shared-memory of the GPU. 
Because access speed from on-chip memory further speeds up the calculations.
We simply apply this procedure to other elements and store this ordered replacements in CRS format using $ia$ and $ja$  arrays \cite{saad2003iterative}. The CRS format is a very popular method for storing neighbor connectivity in unstructured mesh topologies.


\section{Accelerator programming with OpenACC}
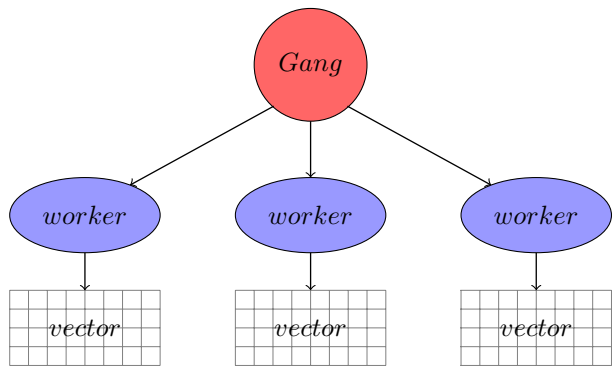
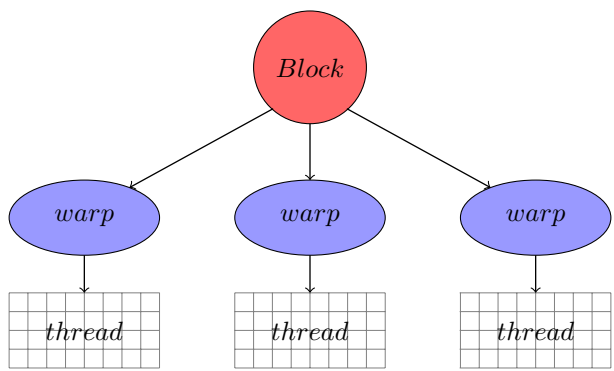
\begin{figure}[h]
\centering
\subfloat[OpenACC]{\begin{tikzpicture}
\filldraw[fill=red!60!white, draw=black]  ( 0, 0)  circle[radius=0.75] node{$Gang$};
\filldraw[fill=blue!40!white, draw=black] (-3,-2)  ellipse (1cm and 0.5cm) node{$worker$};
\filldraw[fill=blue!40!white, draw=black] ( 0,-2)  ellipse (1cm and 0.5cm)  node{$worker$};
\filldraw[fill=blue!40!white, draw=black] ( 3,-2)  ellipse (1cm and 0.5cm)  node{$worker$};
\draw[step=0.25cm,gray,very thin] (-4,-4) grid (-2,-3);
\draw[step=0.25cm,gray,very thin] (-1,-4) grid (1,-3);
\draw[step=0.25cm,gray,very thin] ( 1.99,-4) grid (4,-3);
\draw[draw=none] ( -3.0, -3.5) circle[radius=0.2] node{$vector$};
\draw[draw=none] ( 0, -3.5) circle[radius=0.2] node{$vector$};
\draw[draw=none] ( 3, -3.5) circle[radius=0.2] node{$vector$};
\draw[->,line width=0.5pt] (0.49,-0.55) -- (2.4,-1.6);
\draw[->,line width=0.5pt] (-0.49,-0.55) -- (-2.4,-1.6);
\draw[->,line width=0.5pt] (0,-.75) -- (0,-1.5);
\draw[->,line width=0.5pt] (-3,-2.5) -- (-3,-3.0);
\draw[->,line width=0.5pt] (0,-2.5) -- (0,-3.0);
\draw[->,line width=0.5pt] (3, -2.5) -- ( 3,-3.0);
\end{tikzpicture}
\label{fig:openACCParallelism}
}

\subfloat[OpenACC mapped to an NVDIA GPU]{\begin{tikzpicture}
\filldraw[fill=red!60!white, draw=black]  ( 0, 0)  circle[radius=0.75] node{$Block$};
\filldraw[fill=blue!40!white, draw=black] (-3,-2)  ellipse (1cm and 0.5cm) node{$warp$};
\filldraw[fill=blue!40!white, draw=black] ( 0,-2)  ellipse (1cm and 0.5cm)  node{$warp$};
\filldraw[fill=blue!40!white, draw=black] ( 3,-2)  ellipse (1cm and 0.5cm)  node{$warp$};
\draw[step=0.25cm,gray,very thin] (-4,-4) grid (-2,-3);
\draw[step=0.25cm,gray,very thin] (-1,-4) grid (1,-3);
\draw[step=0.25cm,gray,very thin] (1.99,-4) grid  (4,-3);
\draw[draw=none] ( -3.0, -3.5) circle[radius=0.2] node{$thread$};
\draw[draw=none] ( 0, -3.5) circle[radius=0.2] node{$thread$};
\draw[draw=none] ( 3, -3.5) circle[radius=0.2] node{$thread$};
\draw[->,line width=0.5pt] (0.49,-0.55) -- (2.4,-1.6);
\draw[->,line width=0.5pt] (-0.49,-0.55) -- (-2.4,-1.6);
\draw[->,line width=0.5pt] (0,-.75) -- (0,-1.5);
\draw[->,line width=0.5pt] (-3,-2.5) -- (-3,-3.0);
\draw[->,line width=0.5pt] (0,-2.5) -- (0,-3.0);
\draw[->,line width=0.5pt] (3, -2.5) -- ( 3,-3.0);
\end{tikzpicture}}
\caption{Levels of parallelism for accelerators}
\label{fig:gang}
\end{figure}

Portability and maintenance are becoming increasingly important in the realm of the scientific software development.
Ideally, a portable software is hardware independent and therefore, allows for switching between platforms with different architectures with ease. 
While the hardware trend in large clusters is to use GPUs, there are still a large number of clusters that rely on CPUs only, or provide accelerators from other vendors. 
Apart from differences in the hardware, even within the same architectural family, new designs often differ significantly from their predecessor. 
To address these issues, \texttt{OpenACC} standard \cite{openacc2011openacc} 
was introduced in 2011. 
\texttt{OpenACC} is a a portable parallel programming model designed to alleviate labor-intensive accelerator code development without sacrificing portability.
\texttt{OpenACC} assumes that the accelerator compartmentalizes multiple processing elements (PE) that can execute a given instruction concurrently. Each PE is assumed to be capable of performing vector operations. The goal is to abstract the programming model to every architecture that compartmentalizes PEs.  \texttt{OpenACC} provides three levels of parallelism: \textit{gang, worker} and \textit{vector} as illustrated in Fig. \ref{fig:openACCParallelism}. 

The terminology used in \texttt{OpenACC} is to generalize the programming model for accelerators from any vendor. For example, on NVIDIA GPUs, PE refers to streaming multiprocessors (SM). 
Concepts of \textit{gang, worker} and \textit{vector} map to \textit{block, warp} and \textit{thread} in NVIDIA CUDA terminology, respectively. This correspondence is illustrated in \ref{fig:gang}.
The relation between CUDA and \texttt{OpenACC} can be viewed as similar to the relation Posix Threads has with \texttt{OpenMP}. 
Both \texttt{OpenMP} and \texttt{OpenACC} use compiler directives to generate kernels (or Pthreads) to abstract low-level implementation details away from the developer. 
The first question that comes to mind is that how much performance is lost by replacing low-level CUDA programming model with high-level \texttt{OpenACC} directives? Several studies have been carried out to quantify this possible degradation in performance.
Normat et al. \cite{norman2015case} compare the performance of CUDA Fortran and \texttt{OpenACC} directives for an atmospheric climate kernel. The minor performance loss in using \texttt{OpenACC} outweighs the software effort spent in porting an existing application to CUDA. Besides, \texttt{OpenACC} has been evolving to accommodate new features and improve its performance. For these reasons, we chose to use \texttt{OpenACC} directives to generate kernels in \texttt{PittPack}. \texttt{PittPack} uses PGI's implementation of the \texttt{OpenACC} compiler and targets NVIDIA GPUs. GNU C++ compiler has started supporting \texttt{OpenACC} directives since version 5.1.

\subsection{Streams and Concurrency}
GPUs excel for massively parallel applications, and support task parallelism through the concept of streams.
CUDA streams are supported in \texttt{OpenACC} using the \texttt{async} command. Every issued $async$ command is assigned to a separate stream, which implies that the work will be launched and executed asynchronously in queue as soon as resources become available. 
Stream also enables us to perform asynchronous copy operations to and from the GPU.
Not only does it enable overlapping the inter-nodal \texttt{MPI} communications and data copy to GPUs but also it helps solve the complex tridiagonal system simultaneously by prescribing the real and imaginary parts to different streams.

\begin{figure}[h!]
    \centering
    \includegraphics[scale=0.5]{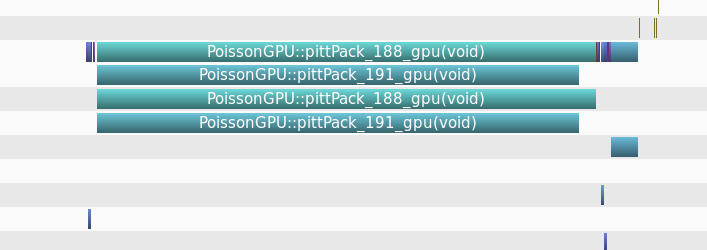}
    \caption{Using multiple streams to achieve concurrency on GPU}
    \label{fig:concurrency}
\end{figure}

Concurrent operations in \texttt{PittPack} is demonstrated on a GPU in Fig. \ref{fig:concurrency}. 
As can be seen from the output of the \texttt{PGI} profiler (\texttt{pgprof}), the two kernels at lines 188 and 191 are launched at the same time.

\subsection{GPU to CPU Binding}
In \texttt{PittPack}, we use a one-to-one mapping between a physical CPU to a GPU. This type of mapping is common in scientific computing on heterogeneous clusters \cite{jacobsen2013multi}.
To accomplish CPU-GPU binding in \texttt{OpenACC}, we ensure that the number of MPI tasks at every node is equal to the number of devices.
\texttt{PittPack} extracts the node number as integer from the $processor\_name$, which is a string. Then it groups the processors that have the same node number and finally creates a communicator for each node.
For one-to-one binding, the size of this communicator should be equal to the number of devices installed on that node. 
If the two numbers do not match an exception is thrown and solution is aborted. 

\section{Well-posedness}
Boundary conditions affect the well-posedness of the boundary value problem for elliptic PDE's.
To illustrate this concept lets take a look at a problem with all Neumann boundary condition for an arbitrary source term.
The Poisson equation with all Neumann boundary  conditions is summarized as follows,
\begin{equation}
\begin{split}
\mathcal{L} u=f \hspace{0.2cm} & \hspace{0.2cm} in \hspace{0.2cm} \Omega,  \\
\hspace{0.3cm} \frac{\partial u}{\partial n}=0 \hspace{0.2cm} & \hspace{0.2cm} on \hspace{0.1cm} \partial \Omega
\end{split}
\label{equ:all_neumann}
\end{equation}
\noindent
Recasting equation \ref{equ:all_neumann} in the (Galerkin) weak form,
\begin{equation}
\begin{aligned}
\int{(\omega \Delta u) d\Omega}=\int{(\omega f) d\Omega}
\end{aligned}
\end{equation}
integrating by parts yields,
\begin{equation}
\begin{aligned}
\int{( \nabla \omega . \nabla u) d\Omega}-\int{(\omega \nabla u . n) ds} =\int{(\omega f) d\Omega}
\end{aligned}
\label{equ:weak}
\end{equation}
This equation should hold for any weighting (test) function depending on the order of the basis polynomials. 
The second term on the left hand side is given as the 
boundary condition in the definition of PDE. 
Assuming $\omega=C$, where $C$ is a constant.
Plugging this in the equation \ref{equ:weak} will result in the following
\begin{equation}
\begin{aligned}
\int{(\nabla u . n) ds} +\int{(f) d\Omega}=0
\end{aligned}
\label{equ:weakFirstOrder}
\end{equation}
This means that the source term can not be arbitrarily specified and it has to follow the condition given in Equation \ref{equ:weakFirstOrder}.
This inconsistency will be translated to the numerical solution procedure in the form of a singular coefficient matrix in one of the tridiagonal systems. By default, this singularity is addressed via keeping the 
corner value which is the mean of the function at z=0 plane fixed. Alternatively, it is also possible to integrate the source term over the volume and subtract it from the source term before proceeding to solve.  

\section{Verification and Validation}
\begin{figure}[h!]
    \centering
    \includegraphics[width=10 cm,height=8cm]{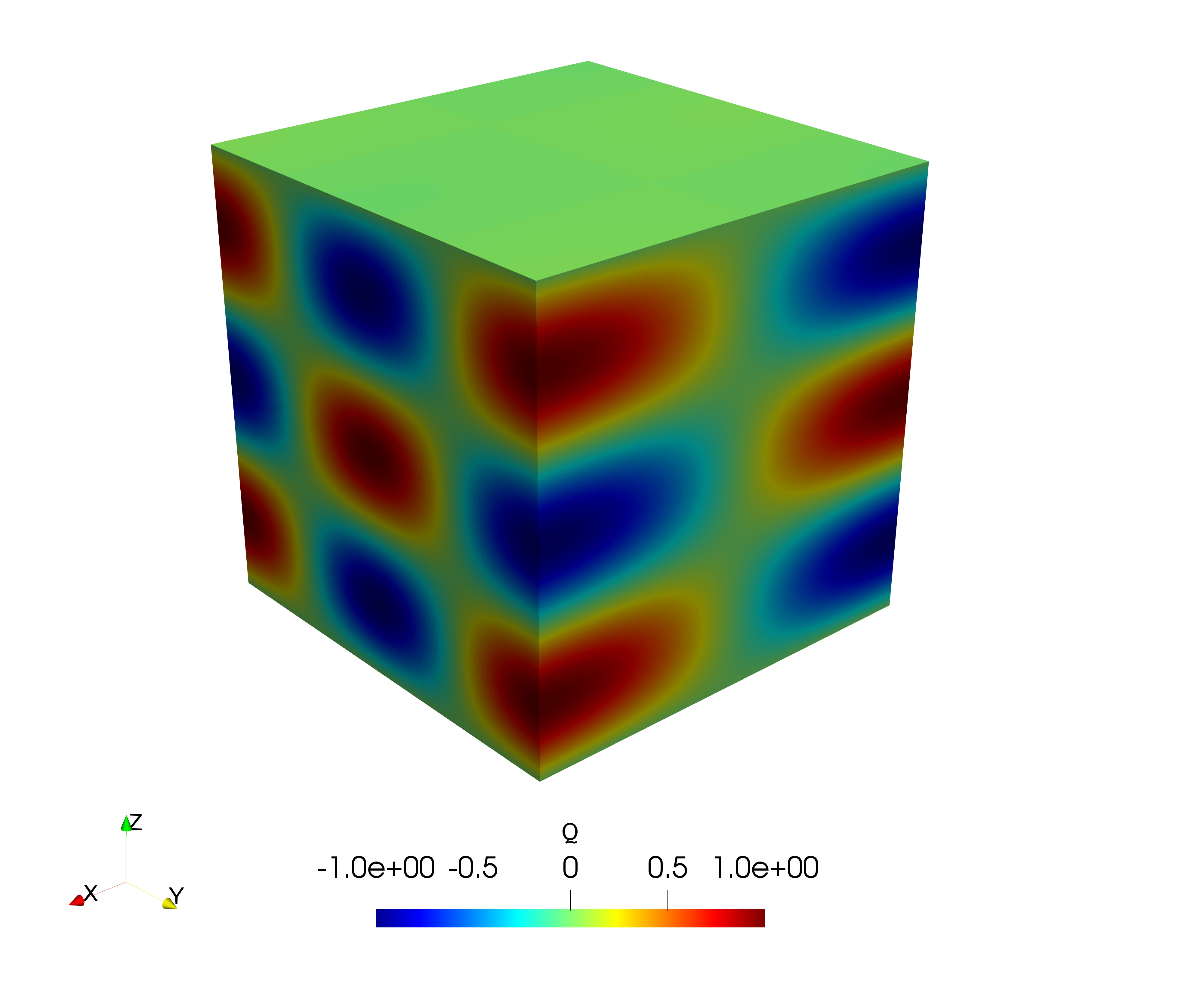}
    \caption{Sample Poisson Solve for NN-NN-DD boundary conditions, grid size $256^{3}$ with 4 GPUs for $\omega_x=1, \omega_y=2$ and $\omega_z=3$}
    \label{fig:samplePoisson}
\end{figure}

The solution of the Poisson’s equation with four NVIDIA Tesla V100 GPUs
are presented in Fig. \ref{fig:samplePoisson}.  
The boundary condition for this case is NN-NN-DD,
in x, y and z-directions, respectively. The exact solution for this test case is $u=cos(\omega_x x) cos(\omega_y y) sin(\omega_z z)$ and the corresponding source term is $f=-(\omega_x^{2}+\omega_y^{2}+\omega_z^{2})*u$.

\subsection{Order of Accuracy}
We use the Method of Manufactured Solutions (MMS) \cite{salari2000code, roache2002code} to verify the order of accuracy of the numerical procedure. Depending on the boundary conditions on each direction appropriate trigonometric function is selected,  while for the periodic case the complex exponential function is used.
The final solution is obtained by tensor product of these functions in $i,j$ and $k$ directions.
This function is plugged into the $\mathcal{L}$ operator. Naturally, it will not satisfy the Laplacian operator and hence will generate source terms. In other words, trying to  construct manufactured solution for Laplace's equation will lead to Poisson's equation. Mathematically speaking, constructing a manufactured solution for Laplace's equation is equivalent to constructing the exact solution for Poisson's equation. We remind here that for a general nonlinear PDEs, such as the Euler equations, this is not the case. In general, MMS can be used to verify order of accuracy. For validation purposes comparison against an exact solution should be considered \cite{Oberkampf2002}. 
Order of accuracy can be defined using the relation $e=C h^{p}$, where $e$ is the error in the numerical solution, $h$ is the mesh spacing, $C$ is a constant and $p$ is the order of the accuracy of the numerical scheme. From this relation, the order of accuracy $p$ can easily be derived using two meshes with spatial resolutions $h_1$ and $h_2$ 
as follows,
\begin{equation}
    p=\frac{log(\frac{e_{2}}{e_{1}})}{log(\frac{h_{2}}{h_{1}}).}
\end{equation}
\noindent
Tables \ref{tab:tb3}- \ref{tab:tb7} show the verification of the order of accuracy given different frequencies for various combination of the boundary conditions. Four NVIDIA V100 GPUs are used for the analysis.   

\begin{table}[h!]
\begin{center}
 \begin{tabular}{c|c|c} 
   $\norm{e}_{2}$    & DD-DD-DD & Order of Accuracy  \\ [0.5ex] 
 \hline 
 N = 128   &  2.30811E-4   &  - \\ 
 \hline 
 N = 256   & 5.76868E-05  &  2.00\\ 
 \hline
 N = 512   &  1.44207E-05 &  2.00\\
 \hline
 N = 1024  &  3.60512E-06  & 2.00\\
  \end{tabular}
 \caption{Verification of order of accuracy using $l_{2}-norm$ of the solution error for DD-DD-DD boundary condition using 4 NVIDIA V100 GPUs with $\omega_x=3$, $\omega_y=4$, $\omega_z=2$}
 \label{tab:tb3}
\end{center}
\end{table}

\begin{table}[h!]
 
\begin{center}
 \begin{tabular}{c|c|c} 
   $\norm{e}_{2}$    &  NN-NN-DD & Order of Accuracy  \\ [0.5ex] 
 \hline
 N = 128   &  1.24261E-04  &   -\\  
 \hline
 N = 256   &  3.10608E-05  &  2.00 \\  
 \hline
 N = 512   &  7.76492E-06  & 2.00 \\ 
 \hline
 N = 1024  &  1.94121E-06  &  1.99999\\ 
  \end{tabular}
 \caption{Verification of order of accuracy using $l_{2}-norm$ of the solution error for NN-NN-DD boundary condition using 4 NVIDIA V100 GPUs,
  with $\omega_x=1$, $\omega_y=2$, $\omega_z=3$}
  \label{tab:tb4}
\end{center}
\end{table}

\begin{table}[h!]
\begin{center}
 \begin{tabular}{c|c|c} 
   $\norm{e}_{2}$    & PP-PP-DD & Order of Accuracy  \\ [0.5ex]
   \hline 
 N = 128   &   1.27925E-03   &   -\\  
 \hline
 N = 256   &   3.19554E-04   &   2.00162\\  
 \hline
 N = 512   &   7.98722E-05  &  2.00028\\ 
 \hline
 N = 1024  &   1.9967E-05  &  2.00225\\ 
  \end{tabular}
 \caption{Verification of order of accuracy using $l_{2}-norm$ for PP-PP-DD boundary condition using 4 NVIDIA V100 GPUs,
  with $\omega_x=5$, $\omega_y=6$, $\omega_z=7$}
  \label{tab:tb5} 
\end{center}
\end{table}

\begin{table}[h!]
\begin{center}
 \begin{tabular}{c|c|c} 
   $\norm{e}_{2}$    & NN-NN-NN & Order of Accuracy  \\ [0.5ex]
   \hline 
 N = 128   &    5.32106E-04 &   -\\  
 \hline
 N = 256   &    1.32945E-04  &   2.0006\\  
 \hline
 N = 512   &   3.32313E-05  &  2.0001\\ 
 \hline
 N = 1024  &   8.3075E-06  &  2.00 \\ 
  \end{tabular}
 \caption{Verification of order of accuracy using $l_{2}-norm$ for NN-NN-NN boundary condition using 4 NVIDIA V100 GPUs, with $\omega_x=1$, $\omega_y=3$, $\omega_z=6$}
 \label{tab:tb6} 
\end{center}
\end{table}

\begin{table}[h!]
\begin{center}
 \begin{tabular}{c|c|c} 
   $\norm{e}_{2}$    & PP-PP-PP & Order of Accuracy  \\ [0.5ex]
   \hline 
 N = 128   &   2.00822E-04 &   -\\  
 \hline
 N = 256   &   5.02009E-05  &   2.0001 \\  
 \hline
 N = 512   &   1.25499E-05  &  2.00 \\ 
 \hline
 N = 1024  &   3.13747E-06  &  2.00 \\ 
  \end{tabular}
 \caption{Verification of order of accuracy using $l_{2}-norm$ for PP-PP-PP boundary condition using 4 NVIDIA V100 GPUs, with $\omega_x=2$, $\omega_y=2$, $\omega_z=2$}
 \label{tab:tb7} 
\end{center}
\end{table}


\section{Parallel Performance Analysis}
We use the Titan supercomputer at the Oak Ridge National Laboratory to assess the scalability of \texttt{PittPack} using up to 16384 GPUs. Note that the Titan is a Cray XK7 system and has a total $18,688$ compute-nodes equipped with an equal number of GPUs. Each node accomodates an AMD-petron with 16 cores. Titan's theoretical peak performance is 27 petaFLOPS. It sustained 17.59 petaFLOPS on the \texttt{LINPACK} benchmark case \cite{Bland2012}. Each GPU on Titan is an NVIDIA Tesla K20X. The nodes in Titan are connected via a three-dimensional torus network.
Each GPU is connected via PCI express 2.0 interface. 

We used the PGI compiler with the flags \texttt{-ta=pinned -fast} to enable GPU timing in \texttt{OpenACC}. Compiler flag \texttt{pinned} enables pinned memory allocation which is essential for asynchronous memory copies between CPU and the GPU. Use of pinned memory enhances data throughput. Note that the operating system (OS) limits the amount of pinned memory allocation since pinned memory has to be physically in the DRAM and only then the RDMA copy operations can be performed without the need for any operations by CPU.  
The flag \texttt{fast} enables the highest level of compiler optimization.
We do not recommend using a unified memory flag as in some cases it degrades the performance.  This is to some extent expected since we rely on the compiler to decide when to copy back and forth from the GPU.

Another important point here is that in the weak scaling, we use the timing for $np=4$ as our reference time rather $np=1$. The reason for this is two fold.  First, if we use the same code with $np=1$, the copy operations to and from GPU are all done in sequence and hence there is no overlap. If we use this value as the reference time the parallel effeciency will be over estimated.
Second option is to use a single GPU version of the \texttt{PittPack}, the run time for this code is actually very low. This is due to the different design as now there is no need for transpose and data shuffling. FFT transforms may be carried out with stride using \texttt{CuFFT} library. This will drastically degrade scaling analysis.  Since numerous routines do not get invoked and hence, this gives a much better time to solution. 

\subsection{Weak Scaling Analysis of Pinned Memory Runs}
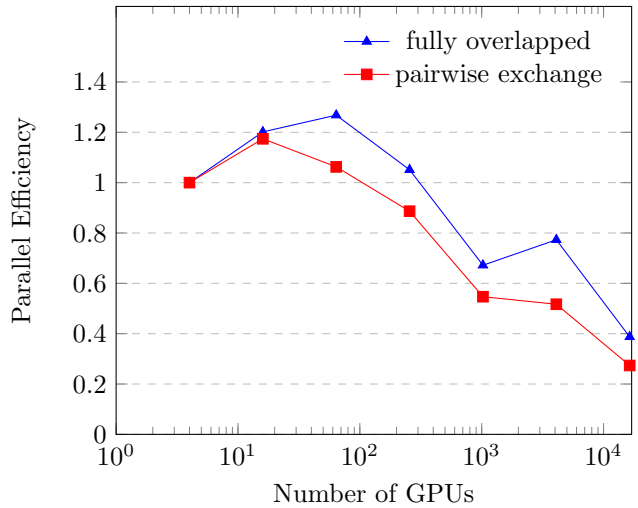
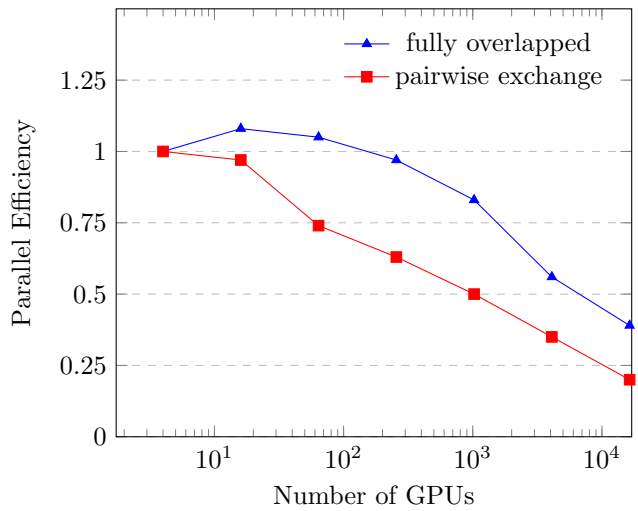
\begin{figure}[h!]
\centering
\subfloat[\texttt{nxChunk=48, nyChunk=48, nzChunk=8192} per GPU, pinned memory]
{
	\centering
	\begin{tikzpicture}
	\begin{semilogxaxis}[
    xlabel={Number of GPUs},
	ylabel={Parallel Efficiency},
	xmin=1, xmax=17000,
	ymin=0, ymax=1.7,
	legend style={draw=none},
	ytick={0,0.2,0.4,0.6,0.8,1.0,1.2,1.4},
	legend pos=north east,
	ymajorgrids=true,
	grid style=dashed,
	]
	\addplot[
	color=blue,
	mark= triangle*,
	]
    coordinates {
	
	(4,5.09/5.09)(16,5.09/4.236)
	(64,5.09/4.014)(256,5.09/4.842)(1024,5.09/7.576)
	(4096,5.09/6.584)(16384,5.09/13.14)
	};
	\addlegendentry{fully overlapped}
	\addplot[
	color=red,
	mark= square*,
	]
    coordinates {
	(4,5.19/5.19)(16,5.19/4.42)
	(64,5.19/4.884)(256,5.19/5.854)(1024,5.19/9.49)
	(4096,5.19/10.04)(16384,5.19/18.97)
	};
	\addlegendentry{pairwise exchange}
	\end{semilogxaxis}
	\end{tikzpicture}
		\label{fig:weakScalingPinned0}
}	

\subfloat[ \texttt{nxChunk=64, nyChunk=64, nzChunk=16384} per GPU, pageable memory]
{
	\centering
	\begin{tikzpicture}
	\begin{semilogxaxis}[
    xlabel={Number of GPUs},
	ylabel={Parallel Efficiency},
	xmin=0, xmax=17000,
	ymin=0, ymax=1.5,
	ytick={0,0.25,0.5,0.75,1.0,1.25},
	legend style={draw=none},
	legend pos=north east,
	ymajorgrids=true,
	grid style=dashed,
	]
	\addplot[
	color=blue,
	mark= triangle*,
	]
    coordinates {
	
	(4,1.0)(16,1.08)
	(64,1.05)(256,0.97)(1024,0.83)
	(4096,0.56)(16384,0.39)

	
	};
	\addlegendentry{fully overlapped}
	\addplot[
	color=red,
	mark= square*,
	]
    coordinates {
	
	(4,1)(16,.97)
	(64,.74)(256,.63)(1024,.50)
	(4096,.35)(16384,0.20)

	};
	\addlegendentry{pairwise exchange}

	\end{semilogxaxis}
	\end{tikzpicture}
	\label{fig:weakScalingPageable}
}
 \caption{Weak scaling analysis of \texttt{PittPack} performed upto 16384 GPUs on Titan}
	
\end{figure}

In this section, we report the weak scaling analysis of \texttt{PittPack}. In weak scaling the amount of computation per MPI process is kept constant by increasing the computational problem size linearly with the number of parallel processes, which is the number of GPUs in our case.
The first test case uses a mesh size of $48\times48\times8192$ per GPU, which is approximately  $309.2$ billion mesh points using 16384 GPUs. 
Figure \ref{fig:weakScalingPinned0} compares the weak scaling performance of two communication patterns available in \texttt{PittPack}. Note here, that the K20 GPUs only have two asynchronous engines. As mentioned earlier we expect the fully overlap version to perform better, and the performance analysis conforms with the theoretical analysis
albeit using much more memory.

\subsection{Weak Scaling Analysis of Pageable Memory Runs}
Pinned memory not only helps the asynchronous copies but also improves the throughput of data transfer between CPU and GPU.
However, OS restricts the amount of pinned memory allocations. Therefore, in order to compute the solution on a mesh with more than trillion point, we have to revert back to the unpinned memory case. Fortunately, this is easily accomplished in \texttt{OpenACC} by removing the compiler flag \texttt{pinned}.

The second test case uses a maximum mesh size of $64\times64\times16384$ per GPU, which amounts to approximately $1.1$ trillion mesh points on 16384 GPUs. Figure \ref{fig:weakScalingPageable} represents the weak scaling for the fully overlapped versus pairwise exchange nodal communication patterns this time with pageable memory allocation.
In addition to the parallel efficiency graphs, the mesh size versus the execution time in seconds are provided for the pairwise exchange method used in this study in Table \ref{tab:WeakRunTime}. Each execution time reported in the table is an average of five separate runs, except for the 16384 GPU case, which is the average of only two separate runs because of the difficulties in accessing nearly the entire supercomputer. This is due to the fact that (logical) processes to (physical) processor mapping for every execution is not unique and it is determined by the job scheduler based on availability of the nodes. For a leadership cluster with thousands of nodes this variation is even more pronounced when partially utilized.

\begin{table}[h!]
\label{table:3} 
\begin{center}
 \begin{tabular}{c|c|c|c|c|c} 
     \multicolumn{2}{c}{} & \multicolumn{2}{|c|}{Fully-Overlapped} & \multicolumn{2}{c}{Pairwise-Exchange}     \\ [0.5ex] 
 \hline
   nGPUs    &  Mesh Size & Time (s)   & $\epsilon$($\%$)  & Time(s) & $\epsilon$ ($\%$)  \\ [0.5ex] 
 \hline
  4   &   2.6844E08  & 14.86  &  101  & 13.57 & 100  \\  
 \hline
  16   &   1.0737E09 & 13.64  &  108  & 13.98 & 97 \\ 
 \hline
  64  &    4.2950E09 & 14.06  &  105 & 18.27 &  74 \\ 
 \hline
  256  &   1.7180E10 & 15.23  &  97 & 21.53 &  63\\ 
 \hline
  1024  &   6.8719E10 & 17.98  &  82 & 26.74  &  50\\ 
 \hline
  4096  &   2.74878E11 & 26.46  &  56 &  39.57 &  35\\ 
 \hline
  16384  &  1.09951E12 & 36.36  &  38 &  66.49 & 20\\ 
  \end{tabular}
 \caption{Weak scaling analysis up to an extreme scale mesh with more than a trillion computational points}
 \label{tab:WeakRunTime}
\end{center}
\end{table}

\subsection{Comparison with Other Software}
\texttt{AFiD-GPU} \cite{zhu2018afid} is an open-source incompressible Navier-Stokes solver developed with a mixed MPI-CUDA Fortran implementation. \texttt{AFiD} simulates wall-bounded buoyancy-driven turbulent flows.
We compiled and executed \texttt{AFiD} on Titan and record the execution time of its pressure Poisson solver and compare it with the performance of \texttt{PittPack}. The numerical methods used to discretize and solve the Poisson's equations are essentially the same.
The number of GPUs used for this test case is 1024. Here, we only time the fully-overlapped communication pattern in \texttt{PittPack} with pinned memory.
\begin{table}[h!]
\begin{center}
 \begin{tabular}{c|c|c|c} 
   mesh size    & \texttt{AFiD} & \texttt{PittPack} & ratio\\ [0.5ex]
   \hline
 Nx=Ny=Nz = 2048  & 1.35 (s)  &  1.95 (s)   &   1.44 \\  
  \end{tabular}
\end{center}
 \caption{ \texttt{AFid} versus \texttt{PittPack} using 1024 GPUs}
 \label{tab:compareAfid}
\end{table}
\noindent
There are several studies that aim at comparing compiler based directive approach versus low-level programming.
In this connection, Norman et al. \cite{norman2015case} report the speed up of the factor [1.35-1.55] for using \texttt{CUDA FORTRAN} as opposed to \texttt{OpenACC FORTRAN} directives. The ratio obtained in this study is as given in Table \ref{tab:compareAfid} is 1.44.  This value belongs to the interval reported in \cite{norman2015case}.
Additionally, \texttt{PittPack} is an object-oriented software developed using the \texttt{C++} programming language so additional overhead is also expected.  

\subsection{GPU vs. CPU Performance Comparison}
We compare the  \texttt{MPI}-only CPU performance against the hybrid \texttt{MPI-OpenACC} GPU execution. To perform a fair comparison, we compare the entire CPU resources on a single node against the single Nvidia Tesla K20x GPU available on Titan supercomputer.
This means that for each run, the number of MPI tasks for the MPI-only CPU execution is 16 times more than the number of MPI tasks for the \texttt{MPI-OpenACC} GPU execution.
The communication pattern for these two choices are not exactly the same. However, the two approaches do have the same hops over the network, considering that 16 of those extra communications are intra-nodal in the MPI-only case.

\begin{table}[h!]
\label{table:3} 
\begin{center}
 \begin{tabular}{c|c|c|c|c|c} 
   Number of Nodes    &  GPUs & CPUs & GPU time & CPU time & Speedup  \\ [0.5ex] 
 \hline
  1   &   1  & 16 &       5.09  &   7.73 & 1.51 \\  
 \hline
  4   &   4  & 64 &       5.09  &   5.81 &  1.14 \\  
 \hline
  16   &   16 & 256 &     4.23 &    5.21 &   1.23 \\ 
 \hline
  64  &    64 & 1024 &    4.01 & 5.13    & 1.28\\ 
 \hline
  256  &   256 & 4096 &   4.84 & 7.01     & 1.45 \\ 
 \hline
  1024  & 1024 & 16384 &  7.58 & 9.15     & 1.21 \\ 
 \hline
  4096 & 4096 & 65536 &   6.58  & 9.26    & 1.41 \\ 
 \hline
  16384  &  16384 & 262144 & 13.14  & 14.76 & 1.12 \\ 
  \end{tabular}
 \caption{GPU versus CPU performance analysis with \texttt{nxChunk=48, nyChunk=48, nzChunk=2048} per node}
 \label{tab:CompareChips}
\end{center}
\end{table}
\noindent
The direct solution algorithm for Poisson's equation is a communication intensive algorithm .Therefore, the results heavily depend on the  the PCI bandwidth as well as the network topology. Titan uses PCIe 2.0 interface. It uses three-dimensional torus for its interconnect topology. 
Additionally, the batch parallelism in the problem increases as the mesh dimensions in x, and y direction increase. We do not show the best case scenario as it would be too optimistic. 
The test case set up in the Table \ref{tab:CompareChips} favors the CPU implementation since it requires less nodal communications per task. Despite all these factors GPU solution is still faster by 12 percent when the highest number of nodes are utilized.   

It is also important to quantify the performance enhancement in modern GPU architectures. To demonstrate this, we report the run-times using different GPU architectures. For this study we use K40, V100 GPUs. To eliminate the effects of the network only single GPU is used for this study.
We use the run-time of the K40 as the reference time to calculate the speed-up.

\begin{table}[h!]
\label{table:3} 
\begin{center}
 \begin{tabular}{c|c|c} 
  GPUs &   time(s) & Speed-up  \\ [0.5ex] 
 \hline
  K40 &   85.86  &  1.0 \\  
   \hline
  V100 &   32.84  & 2.62  \\  
  \end{tabular}
 \caption{Performance of \texttt{PittPack} using different GPU architectures for a mesh with \texttt{nxChunk=512, nyChunk=512, nzChunk=512} elements}
 \label{tab:CompareChips}
\end{center}
\end{table}

\section{General Remarks on PittPack}
The main functionalities for the pencil decomposition and nodal communications are contained in class  \texttt{PencilDcmp}. Classes \texttt{PoissonCPU} and \texttt{PoissonGPU} inherit from this class as demonstrated in Listing \ref{fig:class_hier}. This way each class implements its own specialized set of methods and makes swicthing between CPU and GPU trivial. Preprocessor directives are used to disable unnecessary portions of code depending on the loaded compiler. The configuration file is set up such that it loads all the modules and updates the paths for several national leadership class supercomputers for open science.
The software is mainly designed for clusters with GPU's but it can also be used with regular CPU clusters. 
In our analysis each CPU is bound by a single GPU. We use \texttt{OpenACC} directives and PGI compiler to generate GPU kernels. 
We also note that some of the \texttt{C++} features such as virtual functions, function pointers are not supported with PGI yet. We do not utilize those features of \texttt{C++} for the purpose of portability and ease of use.
\definecolor{zzttqq}{rgb}{0.6,0.2,0}
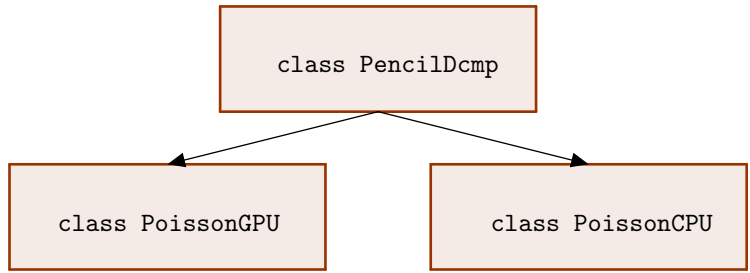
\begin{figure}[h!]
\centering
\begin{tikzpicture}[line cap=round,line join=round,>=triangle 45,x=1cm,y=1cm, scale=0.35]\clip(-14.124575823647707,-11.08619009148831) rectangle (14.117423524433153, 0.03677760599375);\fill[line width=2pt,color=zzttqq,fill=zzttqq,fill opacity=0.10000000149011612] (-6,0) -- (6,0) -- (6,-4) -- (-6,-4) -- cycle;\fill[line width=2pt,color=zzttqq,fill=zzttqq,fill opacity=0.10000000149011612] (-2,-6) -- (-2,-10) -- (-14,-10) -- (-14,-6) -- cycle;\fill[line width=1pt,color=zzttqq,fill=zzttqq,fill opacity=0.10000000149011612] (2,-6) -- (2,-10) -- (14,-10) -- (14,-6) -- cycle;\draw [line width=1pt,color=zzttqq] (-6,0)-- (6,0);\draw [line width=1pt,color=zzttqq] (6,0)-- (6,-4);\draw [line width=1pt,color=zzttqq] (6,-4)-- (-6,-4);\draw [line width=1pt,color=zzttqq] (-6,-4)-- (-6,0);\draw [line width=1pt,color=zzttqq] (-2,-6)-- (-2,-10);\draw [line width=1pt,color=zzttqq] (-2,-10)-- (-14,-10);\draw [line width=1pt,color=zzttqq] (-14,-10)-- (-14,-6);\draw [line width=1pt,color=zzttqq] (-14,-6)-- (-2,-6);\draw [line width=1pt,color=zzttqq] (2,-6)-- (2,-10);\draw [line width=1pt,color=zzttqq] (2,-10)-- (14,-10);\draw [line width=1pt,color=zzttqq] (14,-10)-- (14,-6);\draw [line width=1pt,color=zzttqq] (14,-6)-- (2,-6);\draw [->,line width=0.5pt] (0,-4) -- (-8,-6);\draw [->,line width=0.5pt] (0,-4) -- (8,-6);
\draw (-4.2059868691559898,-1.5099283408767816) node[anchor=north west] {\texttt{class PencilDcmp}};
\draw (-12.51956991150097,-7.559070125060661) node[anchor=north west] {\texttt{class PoissonGPU}};
\draw (3.92250580949422,-7.553108143154292) node[anchor=north west] {\texttt{class PoissonCPU}};
\end{tikzpicture}
\caption{Illustration of the Class Hierarchy for \texttt{PittPack}}
\label{fig:class_hier}
\end{figure}
\noindent

As noted earlier, \texttt{PittPack} utilizes chunked-pencil decomposition strategy. 
Helper class \texttt{chunkedArray} is constructed to abstract the details of this data structure for the user and 
to facilitate access to elements with given chunk id $ID$ and ($i$, $j$, $k$) indices. 
Class \texttt{chunkedArray} accomplishes this mainly by operator overloading. 
This approach consequently improves high-level construction of the communication patterns in class \texttt{PencilDcmp}.
Major functionality of this class is shown in Listing \ref{fig:chunkedArray}.

\definecolor{mygreen}{rgb}{0,0.6,0}
\definecolor{mygray}{rgb}{0.5,0.5,0.5}
\definecolor{mymauve}{rgb}{0.58,0,0.82}
\definecolor{backcolour}{rgb}{0.95,0.95,0.92}
\lstdefinestyle{customc}{
  belowcaptionskip=1\baselineskip, 
  breaklines=true,
  frame=L,
  xleftmargin=\parindent,
  language=C++,
  showstringspaces=false,
  backgroundcolor=\color{backcolour},
  basicstyle=\footnotesize\ttfamily,
  keywordstyle=\bfseries\color{red},
  commentstyle=\itshape\color{black},
  identifierstyle=\color{blue},
  stringstyle=\color{orange},
}
\lstset{style=customc}
\lstset{caption={Class ChunkedArray},label=chunkedArray}
\lstset{basicstyle=\scriptsize}
\begin{center}
\begin{minipage}{\linewidth}
\begin{lstlisting}[frame=single]
class ChunkedArray
{
    private:
    int nChunk;    /*!< number of chunks owned by each processor*/
    int Nx;        /*!< Nx is the number of grid point in X for the whole grid for each process  */
    int Ny;        /*!< Nx is the number of grid point in Y for the whole grid for each process */
    int Nz;        /*!< Nx is the number of grid point in Z for the whole grid for each process */
    int arraySize;  /*!<Total size of array */
    double Xa, Xb, Ya, Yb, Za, Zb; /*!< Block coordinates to be passed on to Class Phdf5 for visualization*/
    double *__restrict__ P = nullptr; /*!< Restricted main pointer tohold data*/
    int chunkSize = 0;
        
    public:
    ChunkedArray() {};       /*!< Constructor  */
    void moveHostToDevice(); /*!< Copys from GPU to CPU  */
    void moveDeviceToHost(); /*!< Copys from GPU to CPU  */
    PittPackResult allocate( int *n, int nbl );
    int size();
    int getChunkSize();
    void setDirection( int dir );
    void setCoords( double *X );
    #pragma acc routine
    inline  double &operator()( int i, int j, int k, int index );
    #pragma acc routine 
    inline double &operator()( int i, int j, int k );
    #pragma acc routine
    inline  double &operator()( int chunkId, int dir, int i, int j, int k, int index );
    ~ChunkedArray(); 
\end{lstlisting}
\end{minipage}
\label{fig:chunkedArray}
\end{center}
For the GPU programing \texttt{PittPack} uses \texttt{PGI} implementation of the \texttt{OpenACC} directives along with NVIDIA's \texttt{cuFFT} library.

\section{Limitations of \texttt{PittPack}}
Following are the expressed limitations of the released software 
\begin{enumerate}
    \item Communicator size must be a square integer.
    \item Mesh size in the $z$-direction for CR-P algorithm should be $2^{n}-1$, where $n$ is a positive integer
    \item \texttt{PittPack} only solves the Poisson' equation on directionally uniform grids. 
    \item Mesh size in the $z$-direction for PCR algorithm should be less than 4096 in $z$-direction
\end{enumerate}
The first limitation is due to the communication pattern design. The second limitation comes from the PCR algorithm and its heavy reliance on  resources per block which depends on the GPU architecture. The first three limitations will be addressed in the future releases of \texttt{PittPack}.

\section{Conclusion}
We have presented a mixed \texttt{MPI-OpenACC} parallel implementation and performance of an open-source software for the direct solution of the Poisson's equation on directionally uniform Cartesian grids. The solver is portable and can execute on CPU-only or GPU-only clusters. The solver uses the FFTW or the cuFFT libraries depending on the execution mode. We considered different tridiagonal solvers to assess their impact on the overall run time. We have found that low-memory Thomas algorithm produces satisfactory results relative to its parallel cyclic reduction alternatives and adopted the Thomas algorithm as the tridiagonal solver in our scaling analysis.
\texttt{PittPack} is designed to take advantage of the GPUs on heterogeneous computing clusters. We have considered two different communication patterns to perform FFT in parallel for extreme-scale computing with more than one trillion mesh points in the computations. An efficient communication pattern is introduced that enhances the scalability of the software at extreme scale by eliminating unnecessary memory usage and exploiting nodal communication and data transfer overlap between hosts and devices. This approach is also network friendly and does not overwhelm the network.
Fully overlapped communication pattern is also included for the cases when available GPU memory is not a limiting issue. Implementation of both of the communication patterns are facilitated using chunked-pencil decomposition strategy.

Using up to 16384 GPUs on a leadership class supercomputer, we have demonstrated that the pairwise exchange algorithm is more suitable for handling large problem sizes, while the fully overlapped communication pattern maximizes execution time. Pairwise exchange algorithm is network friendly at extreme scale since the number of MPI send and receives does not grow linearly with the number of neighbors. The fully-overlapped version can overwhelm the network as problem size gets larger and, therefore, is not recommended for extreme-scale computing. In addition, the number of asynchronous engines on the GPU will also limit the fully overlapped communication pattern. This issue is expected to improve on latest supercomputers with new GPUs with 7 asynchronous engines.
We make both of the methods of communication available in \texttt{PittPack} as a user option. 
\texttt{PittPack} is offered as open source and is freely available on github at {\color{blue}{https://github.com/GEM3D/PittPack}}.

\section*{Acknowledgement}
This material is based upon work supported by the National Science Foundation under Grant No. 1440638 and Army
Research Office under Grant No. W911NF-17-1-0564.
This work was developed and tested on XSEDE platforms \cite{towns2014xsede}. 
This work used the Bridges system \cite{nystrom2015bridges}, which is supported by NSF award number ACI-1445606, at the Pittsburgh Supercomputing Center (PSC). This research used resources of the Oak Ridge Leadership Computing Facility at the Oak Ridge National Laboratory, which is supported by the Office of Science of the U.S. Department of Energy under Contract No. DE-AC05-00OR22725.
The authors would like to thank Matt Colgrove and John Urbanic for their technical support with \texttt{OpenACC}.
We extend our special thanks to Cheng-Nian Xiao for verifying \texttt{PittPack}.

\section*{References}

\appendix
\section*{Appendix: PittPack Poisson's solve example}
The use of the \texttt{PittPack} is illustrated in a simple example.
All the required \texttt{\texttt{MPI}\_Init()} and \texttt{\texttt{MPI}\_Finalize()}, GPU initialization and binding are encapsulated to provide a very user-friendly and simple interface.
\texttt{PittPack} keeps the CPU and GPU objects separately therefore the only modification from
switching from GPU to CPU is at the definition of the object. Thanks to C++'s polymorphism. 

\lstset{caption={Illustration of utilization of PittPack},label=PittPackUtilize}
\lstset{basicstyle=\scriptsize}
\begin{center}
\begin{minipage}{\linewidth}
\begin{lstlisting}[frame=single]
.....

#include "PittPack.h"

int main(int argc, *pArgs[])
{
  int n0=Nx // number of grid points in x direction
  int n1=Ny // number of grid points in y direction
  int n2=Nz // number of grid points in z direction
  
  // construct the Poisson solver object
  PoissonGPU Pois(argc,pArgs, int n0, int n1, int n2) 
  
  // or if you are using CPU use the CPU version
  // PoissonCPU M(argc,pArgs, int n0, int n1, int n2) 
  
  bc={'N','N','P','P','D','D'};
  
  Pois.assignBoundary(bc);
  
  // provide the source term, i.e. right hand side (f), size is the size of array per GPU
  // or per CPU rank, data-movement is implicitly performed
  double *P=new double[size];
  
  Pois.assignRhs(P);
  
  Pois.pittPack(P);
  
  delete [] P;
  return(0)
}
\end{lstlisting}
\end{minipage}
\end{center}

\end{document}